\documentclass[a4paper, 11pt]{article}
\usepackage{graphicx}
\usepackage{amssymb}
\usepackage{epstopdf}
\usepackage{fancyhdr}
\usepackage{amsmath}
\usepackage{amsthm}
\usepackage{hyperref}
\usepackage{makeidx}
\usepackage{mathdesign}
\usepackage{color}
\usepackage{soul}
\usepackage[top=2.2 cm,bottom=2.2 cm,left=2.5 cm, right=2.5 cm]{geometry}

\definecolor{purple}{rgb}{.9,0,.9}

\makeatletter
\let\orgdescriptionlabel\descriptionlabel
\renewcommand*{\descriptionlabel}[1]{%
  \let\orglabel\label
  \let\label\@gobble
  \phantomsection
  \edef\@currentlabel{#1}%
  \let\label\orglabel
  \orgdescriptionlabel{#1}%
}
\makeatother

\newcommand{\Real}{\mathbb{R}}

\newcommand{\rfa}{\qquad {\rm for \ all}\ }

\newcommand{\bb}{{\bf b}}
\newcommand{\be}{{\bf e}}

\newcommand{\bu}{{\bf u}}
\newcommand{\bw}{{\bf w}}
\newcommand{\bA}{{\bf A}}
\newcommand{\bC}{{\bf C}}

\newcommand{\bR}{{\bf R}}

\newcommand{\beqn}{\begin{equation}}
\newcommand{\eeqn}{\end{equation}}

\newcommand{\bbE}{\mathbb{E}}

\newcommand{\bnu}{\boldsymbol{\nu}}
\newcommand{\ve}{\varepsilon}

\def\Xint#1{\mathchoice
{\XXint\displaystyle\textstyle{#1}}%
{\XXint\textstyle\scriptstyle{#1}}%
{\XXint\scriptstyle\scriptscriptstyle{#1}}%
{\XXint\scriptscriptstyle\scriptscriptstyle{#1}}%
\!\int}
\def\XXint#1#2#3{{\setbox0=\hbox{$#1{#2#3}{\int}$ }
\vcenter{\hbox{$#2#3$ }}\kern-.58\wd0}}

\def\dashint{\Xint-}  

\title{The impact of microfibril orientations on the biomechanics of plant cell walls and tissues: modelling and simulations
\thanks{M.~Ptashnyk and B.~Seguin gratefully acknowledge the support of the EPSRC First Grant EP/K036521/1 ``Multiscale modelling and analysis of mechanical properties of plant cells and tissues''. }
}
\author{Mariya Ptashnyk and Brian Seguin}

\begin{document}
\date{}

\maketitle

\begin{abstract}
\noindent It is known that the orientation of cellulose microfibrils within plant cell walls has an important impact on the morphogenesis of plant cells and tissues.  Viewing the shape of a plant cell as a square prism or cylinder with the axis aligning with the primary direction of expansion and growth, the orientation of the microfibrils within the cell wall on the sides of the cell is known.  However, not much is known about their orientation at the ends of the cell.  Here we investigate the impact of the orientation of cellulose microfibrils within a plant cell wall at the ends of the cell by solving the equations of linear elasticity numerically.  Three different scenarios for the orientation of the microfibrils are considered.  The macroscopic elastic properties of the cell wall are obtained using homogenization theory from the microscopic description of the elastic properties of the cell wall microfibrils and wall matrix.  It is found that the orientation of the microfibrils in the upper and lower parts of cell walls do not affect the expansion of the cell in the direction of its axis but do affect its expansion in the lateral directions.  The arrangement of the microfibrils in the upper and lower parts of cell walls  is especially important in  the case of  directed forces acting on plant cell walls and tissues. \\



\end{abstract}

%\tableofcontents

\section{Introduction} 

Knowing the influence of the microscopic molecular interactions and microscopic structure of plant tissues on the mechanical properties, development, and growth of plants is vital for the agriculture and energy sectors.  The mechanical properties of plant tissues are strongly determined by the mechanical properties of the cell walls surrounding plant cells and by the cross-linked pectin network of the middle lamella which joins individual cells together.  Primary cell walls of  plant cells, that are strong so as to resist a high internal hydrostatic pressure (turgor pressure) and flexible to permit growth, consist mainly of oriented cellulose microfibrils, pectin, hemicellulose, proteins, and water. 
%Cells in a plant tissues are connected by a pectin network of  middle lamella.
%One of the components of plant cells that is not present in animal cells is the cell wall, which must be strong to resist a high internal hydrostatic pressure (turgor pressure) and flexible to permit growth.  The primary wall of a plant cell consists mainly of oriented cellulose microfibrils, pectin, hemicellulose, proteins, and water. Covering the primary cell wall is the pectin rich middle lamella which binds together neighboring cells.  %Since the turgor pressure acts isotropically, it is the microstructure of the cell wall which determines the anisotropic expansion of plant cells.
%The mechanical properties of the cell wall are influenced by the microfibrils, hemicellulose, and pectin.
The orientation, length, and high tensile strength of the microfibrils strongly influences the wall's stiffness.  Hemicelluloses form hydrogen bonds with the surface of cellulose microfibrils, which may effect the mechanical strength of the cell wall by creating a microfibril-hemicellulose network \cite{SBBFetal}.  Pectin, once it is de-esterified and  cross-linked with calcium ions, forms a gel within the primary cell wall and middle lamella and is hypothesized to be one of the main regulators of cell wall elasticity \cite{WHH}.

%Pectin, produced by the cell in a highly methylestrified state, can be modified by the enzyme pectin methylesterase (PME), which removes methyl groups by breaking ester bonds.  The de-esterified pectin is able to form calcium-pectin cross-links, which stiffen the cell wall \cite{PBLBetal, PVWVetal, WG, ZMSR}. It is supposed that calcium-pectin cross-linking chemistry is one of the main regulators of cell wall elasticity \cite{WHH}. {\color{blue} Are the details of the chemistry necessary for this paper?}

%For irreversible deformation, the deposition of new wall materials and the loosening of the cell wall through the breaking of the load-bearing microfibril-hemicellulose network and calcium-pectin cross-links are required, see e.g. \cite{Cosgrove05, Schopfer06}.

Since the turgor pressure acts isotropically, it is the microstructure of the cell wall which determines the anisotropic expansion of plant cells.  More specifically, it is the orientation of the cellulose microfibrils that influences the anisotropic expansion of the cell.  Many plant cells, especially cells in plant roots and stem tissues,  have a primary direction of expansion and less expansion takes place in the directions orthogonal to it, see e.g.~\cite{Green,PP}.  It is well-known that cellulose microfibrils are parallel to the sides of primary cell walls and, particularly in young cells, perpendicular to the main direction of expansion \cite{Green99,SWW,IMac,SC}.  For plant cells whose shape can be approximated by a prism or cylinder with the axis aligned with the primary direction of expansion, which is the case for root cells, this means that the microfibrils within the primary cell wall making up the sides of the cell are parallel to the sides and perpendicular to the axis of the cell. However, the orientation of the microfibrils in the cell wall at the ends of the cell does not appear to be known.  Due to the important role that microfibrils play in the mechanical properties and expansion of the cell wall, knowing the orientation of the microfibrils everywhere is of vital importance.

%{\color{red}Viewing a plant cell as roughly having the shape of a prism or cylinder with the axis aligned with the primary direction of expansion, it is known that in young root cells the cellulose microfibrils within the primary cell wall making up the sides of the cell are parallel to the sides and perpendicular to the axis of expansion, see e.g.~\cite{SWW}.  However, not much is known about the orientation of the microfibrils at the ends---that is, the top and bottom---of the cell.  }

In this paper we investigate the affect of the orientation of the cellulose microfibrils in the upper and lower parts of cell walls  on the deformation of the cell walls  and plant tissues using numerical simulations.  Modeling plant cells as a square prisms with rounded edges, we consider a part of a plant tissue represented by  a central cell surrounded by cells on all sides.  The primary cell wall and middle lamella are modeled as linearly elastic materials and on its internal boundary we specify a traction boundary condition corresponding to the turgor pressure.  The cellulose microfibrils are arranged periodically within the cell wall, see e.g.~\cite{Tomasetal}. The  length scale of microfibrils (their diameter and distance between microfibrils) is much smaller than the scale associated with the thickness of the cell wall.  This smaller length scale will be referred to as the microscale, while the scale associated with the dimensions of the cell wall is called the macroscale.  To obtain the elastic properties of the primary cell wall we follow \cite{PS} and use homogenization theory to find an effective (macroscopic)  elasticity tensor that depends on the orientation of the microfibrils on the microscale. 
It was observed experimentally that calcium-pectin cross-links influence mechanical properties of the cell wall matrix and of middle lamella, e.g.\ \cite{WHH}. The affect of the density of the calcium-pectin cross-links on the elastic properties of the cell wall are modeled through the Young's modulus of the isotropic cell wall matrix.  The microfibrils are assumed to be transversely isotropic.  
The effective elasticity tensors for cell walls  are determined from the microscopic description of the mechanical properties of the microfibrils  and cell wall matrix by  solving numerically the corresponding  `unit cell' problems. Then  using the macroscopic elasticity tensor for different  microfibril orientations we solve numerically the equations of linear  elasticity  with different traction boundary conditions.  The affect  of the length of the cell in the direction of its axis on the deformation of the tissue is also investigated by considering two different cell lengths.

We find that different configurations of orientations of microfibrils  in the plane perpendicular to the main axis of cells has little effect on the expansion of the cells in the direction of its axis, however they do affect the expansion of the cell in the orthogonal directions.  In general, we found that the expansion in the directions aligned with the microfibrils is less than the expansion in the directions orthogonal to the microfibrils.  We also found that the expansion of the cell in the direction of its axis is smaller for shorter cells, which is in accord with Hooke's law.   If there are no applied forces and assuming the same turgor pressure in all cells, the expansion in every direction is smaller than in the case of additional forces acting on a plant tissue.  The difference in the turgor pressure in the neighboring cells cause larger deformations in the directions parallel to the ends of cells, but the absolute value of the maximum of the deformation is negligibly affected by the orientation of the microfibrils in the upper and lower regions of the cell walls.  
The arrangement of the microfibrils  in the upper and lower parts of the cell walls  do have an impact on the elastic deformation of a plant tissue in the case where there are external forces or tissue tension in the directions parallel to the upper and lower parts of the cell walls.

The outline of the paper is as follows.  In Section~\ref{sectmodel}  we specify our model  for  plan tissue biomechanics.  We consider the elastic deformation of the primary cell walls joined by middle lamella and the cell-inside is modelled by  prescribing a turgor pressure.  Next, in Section~\ref{sectnum}, the results of our numerical simulations are presented.  The numerical results are discussed in Section~\ref{sectdisc}.

%%%%%%%%%%%%%%

\section{Statement of the model}\label{sectmodel}

In this section we present our model for the elastic deformations  of  a part of a plant tissue consisting of 27 cells  connected by  middle lamella.  This section is divided into three parts: a description of the geometry of the domain, the presentation of the governing equations and boundary conditions, and the specification of the elasticity tensor on the domains representing the different parts of a plant cell wall and middle lamella.

%The presence and orientation of the microfibrils plays a crucial role in our model.  In different regions within the domain the microfibrils are arranged in different directions.  These regions are specified in Section~\ref{sectgeo}.  

%On the sides of a young plant cell, the orientation of the microfibrils are parallel to the sides and perpendicular to the primary direction of expansion.  %However, not much is known about the orientation of the cellulose microfibrils on the top and bottom of the cell.

%The primary cell wall consists mainly of oriented cellulose microfibrils and a matrix made up of pectin, hemicellulose, structural proteins, and water, while the middle lamella is mostly pectin crosslinked with calcium ions.  The orientation of the cellulose microfibrils and the density of pectin-calcium cross-links greatly influences the mechanical properties of the primary cell wall.  On the sides of a young plant cell, the orientation of the microfibrils are parallel to the sides and perpendicular to the direction of growth.  However, not much is known about the orientation of the cellulose microfibrils on the top and bottom of the cell.

\subsection{Geometry}\label{sectgeo}

\begin{figure}
%\hspace{.8in}\includegraphics[height=2.5in]{8celldomainhold}
\hspace{.8in}\includegraphics[height=2in]{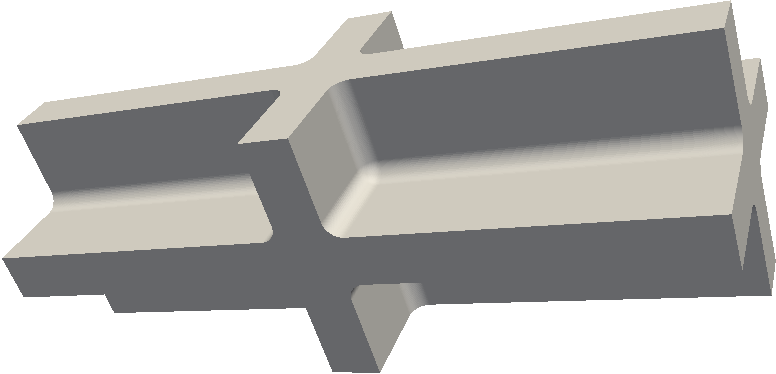}
\put(-318.2,76.7){\rotatebox[origin=c]{4}{$\vector(1,0){28}$}}
%\put(-280,47){\rotatebox[origin=c]{4}{$\vector(1,0){28}$}}
\put(-324,83.8){\rotatebox[origin=c]{40}{$\vector(1,0){28}$}}
%\put(-305,100){\rotatebox[origin=c]{40}{$\vector(1,0){28}$}}
\put(-317.5,60.9){\rotatebox[origin=c]{-60}{$\vector(1,0){28}$}}
%\put(-302,33.8){\rotatebox[origin=c]{-60}{$\vector(1,0){28}$}}
\put(-299,70){$x_3$}
\put(-310,93){$x_2$}
\put(-316,52){$x_1$}
\caption{The domain $\Omega$ consisting of parts of eight cells.  By reflecting this domain over the planes $x_1=0$, $x_2=0$, and $x_3=0$ one obtains a domain that includes the central cell and parts of the 26 cells that surround it.}
\label{8celldom}
\end{figure}

Our geometry is motivated by the structure of cells and tissues in young plant roots.  We assume that the shape of a plant cell can be approximated by a square prism with rounded edges and consider a central cell surrounded by 26 cells, including diagonally adjacent cells.  Choose a $(x_1,x_2,x_3)$ coordinate-system so that the origin is in the center of the central cell, the axes are parallel to the edges of the prism of the central cell, and the $x_3$-axis is aligned with the axis of the central cell.  Moreover, we consider $1$~unit to be $1$ $\mu$m.  We consider the domain $\Omega$ as depicted in Figure~\ref{8celldom}, the bounding box of which is $(0,7.5)\times(0,7.5)\times (0,21.5)$.  By reflecting $\Omega$ over the planes $x_1=0$, $x_2=0$, and $x_3=0$ one obtains a domain that includes the central cell and parts of the $26$ cells that surround it.  The planes $x_1=0$, $x_2=0$, and $x_3=0$ will be called the planes of symmetry.

% A cross-section of $\hat\Omega$ at $x_3=0$ is shown in Figure~\ref{8celldomCS}.

%Consider a plant cell surrounded by cells on all sides, so that a total of $27$ cells are considered.  The axis of the prism is aligned with the primary direction of growth of expansion.  Assume that all of the cells have the same dimensions.  The domain $\hat\Omega$ consists of the region occupied by the primary cell wall and the middle lamella of the central cell and parts of the surrounding cells.  Choose an $(x_1,x_2,x_3)$ coordinate system so that the origin is in the center of the central cell, the axes are parallel to the edges of the prism, and the $x_3$-axis is aligned with the axis of the prism.  Moreover, consider $1$ unit to be $1$ $\mu$m.  A cross-section of $\hat\Omega$ at $x_3=0$ is shown in Figure~\ref{8celldomCS}.

%The domain $\hat\Omega$ is symmetric about the $x_1$, $x_2$, and $x_3$-planes and, as will be specified in greater detail below, so are the governing equations and boundary conditions.  Because of these symmetries, we only consider the domain $\Omega$ depicted in Figure~\ref{8celldom}, which is the part of $\hat\Omega$ in the positive octant.  The bounding box of $\Omega$ is $(0,7.5)\times(0,7.5)\times (0,21.5)$.  

\begin{figure}
\hspace{1.75in}\includegraphics[height=2.5in]{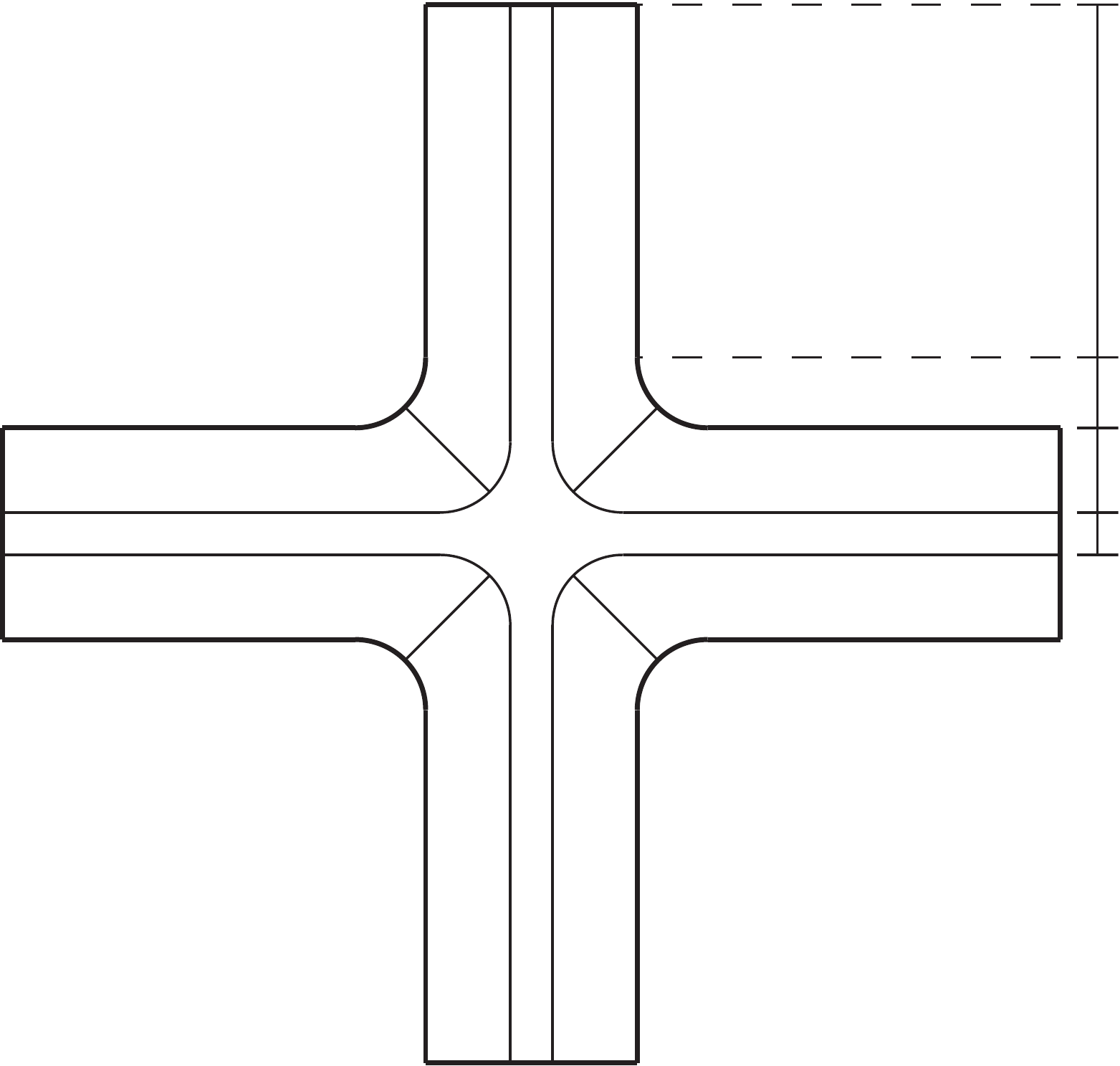}
\put(7,145){$2.5\,\mu$m}
\put(7,98){$0.6\,\mu$m}
\put(7,111){$0.5\,\mu$m}
\put(7,86){$0.3\,\mu$m}
%\put(7,75){$0.6\,\mu$m}
%\put(7,63){$0.6\,\mu$m}
%\put(7,27){$2.5\,\mu$m}
\put(-190,0){\rotatebox[origin=c]{0}{$\vector(1,0){28}$}}
\put(-190,0){\rotatebox[origin=c]{0}{$\vector(0,1){28}$}}
\put(-160,0){$x_1$}
\put(-189,30){$x_2$}
\put(-94,140){\raisebox{.5pt}{\textcircled{\raisebox{-.9pt} {$2$}}}}
\put(-115.5,140){\raisebox{.5pt}{\textcircled{\raisebox{-.9pt} {$2$}}}}
\put(-94,30){\raisebox{.5pt}{\textcircled{\raisebox{-.9pt} {$2$}}}}
\put(-115.5,30){\raisebox{.5pt}{\textcircled{\raisebox{-.9pt} {$2$}}}}
\put(-160,97){\raisebox{.5pt}{\textcircled{\raisebox{-.9pt} {$1$}}}}
\put(-160,77){\raisebox{.5pt}{\textcircled{\raisebox{-.9pt} {$1$}}}}
\put(-50,97){\raisebox{.5pt}{\textcircled{\raisebox{-.9pt} {$1$}}}}
\put(-50,77){\raisebox{.5pt}{\textcircled{\raisebox{-.9pt} {$1$}}}}
%\put(10,0}{\textcircled{1}}
\caption{A cross-section of $\Omega$ at a constant $x_3$-value satisfying $0< x_3< 9.5$ or $12< x_3< 21.5$.  All rounded corners in this figure have a radius of $0.5\,\mu$m.  The regions marked with $1$ have cellulose microfibrils parallel to the $x_1$-axis and regions marked with $2$ have microfibrils parallel to the $x_2$-axis.  The region that is not marked is the middle lamella, which has no microfibrils.  This cross-section is symmetric about the lines $x_1=3.75$, $x_2=3.75$, and $x_1=x_2$.}
\label{XYcrosssect}
\end{figure}

\begin{figure}
\hspace{.4in}\includegraphics[height=2.1in]{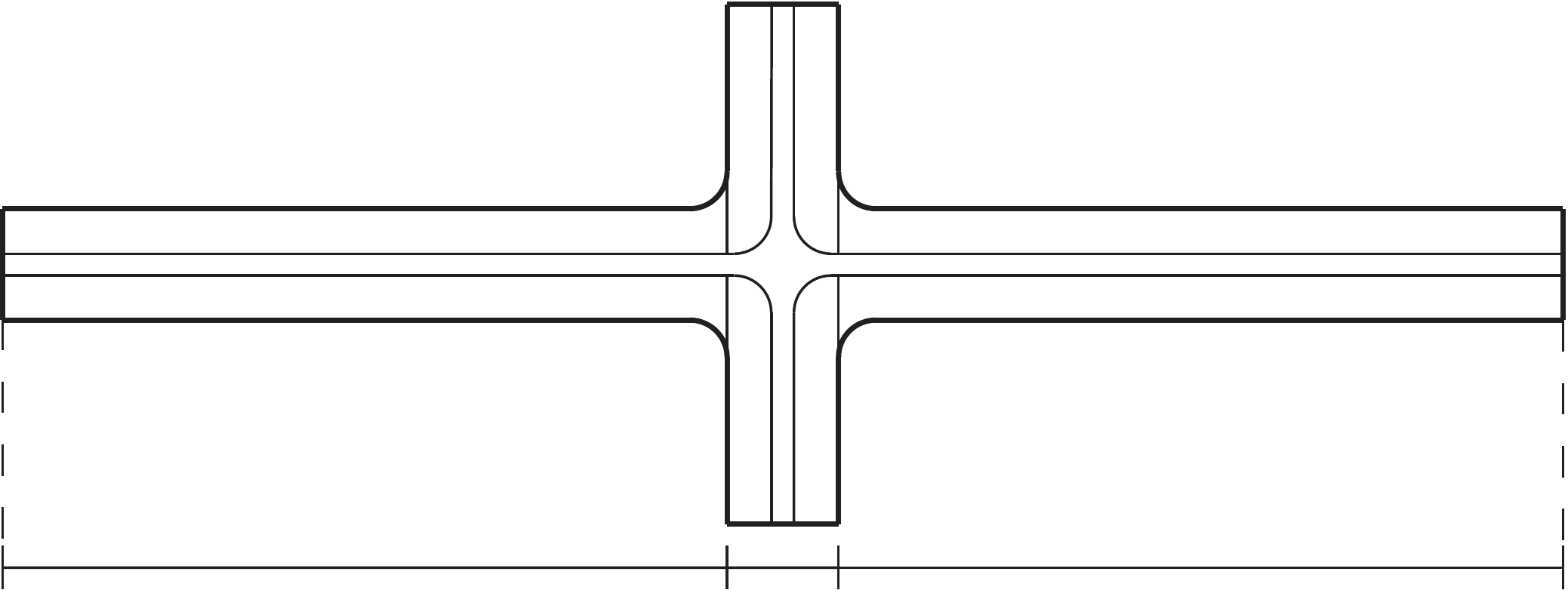}
\put(-315,-10){$10\,\mu$m}
\put(-218,-10){$1.5\,\mu$m}
\put(-100,-10){$10\,\mu$m}
\put(-403,151){\rotatebox[origin=c]{0}{$\vector(1,0){28}$}}
\put(-403,151){\rotatebox[origin=c]{0}{$\vector(0,-1){28}$}}
\put(-403,117){$x_1$}
\put(-375,145){$x_3$}
\put(-95,89){\raisebox{.5pt}{\textcircled{\raisebox{-.9pt} {$2$}}}}
\put(-95,72){\raisebox{.5pt}{\textcircled{\raisebox{-.9pt} {$2$}}}}
\put(-310,89){\raisebox{.5pt}{\textcircled{\raisebox{-.9pt} {$2$}}}}
\put(-310,72){\raisebox{.5pt}{\textcircled{\raisebox{-.9pt} {$2$}}}}
\put(-216,45){\raisebox{.5pt}{\textcircled{\raisebox{0pt} {c}}}}
\put(-199,45){\raisebox{.5pt}{\textcircled{\raisebox{0pt} {c}}}}
\put(-216,120){\raisebox{.5pt}{\textcircled{\raisebox{0pt} {c}}}}
\put(-199,120){\raisebox{.5pt}{\textcircled{\raisebox{0pt} {c}}}}
\caption{A cross-section of $\Omega$ at a constant $x_2$-value satisfying $0< x_2< 2.5$ or $5< x_2< 7.5$.  All of the rounded corners have a radius of $0.5$ $\mu$m.  The regions marked with 2 have cellulose microfibrils parallel to the $x_2$-axis, which is pointed into the page.  The regions marked with c consist of the tops and bottoms of the cells and different microfibril orientations will be considered in these regions.  The region that is not marked is the middle lamella, which has no microfibrils.  This cross-section is symmetric about the lines $x_1=3.75$ and $x_3=10.75$.}
\label{XZcrosssect}
\end{figure}

A cross-section of $\Omega$ at a constant $x_3$-value satisfying $0< x_3 < 9.5$ or $12< x_3 < 21.5$ is shown in Figure~\ref{XYcrosssect}.  The regions with different orientations of the cellulose microfibrils and to the location of the middle lamella in this cross-section are shown in this figure.  The cross-sections of $\Omega$ for a constant $x_3$-value satisfying $9.5<x_3 <12$ are not identical due to the rounded edges of the domain, see Figure~\ref{8celldom}.  A cross-section of $\Omega$ at a constant $x_2$-value satisfying $0< x_2< 2.5$ or $5< x_2< 7.5$ is shown in Figure~\ref{XZcrosssect}.  Once again, the domains with different orientations of the microfibrils and the location of the middle lamella are specified.  Similar to the $x_3$-direction, the cross-sections of $\Omega$ for a constant $x_2$-value satisfying $2.5<x_2<5$ are not identical due to the rounded edges of the domain, see Figure~\ref{8celldom}.  A cross-section of the domain at a constant $x_1$-value satisfying $0< x_1< 2.5$ or $5< x_1< 7.5$ is similar to Figure~\ref{XZcrosssect}.  The thickness of $\Omega$ away from the junctions between two sections of the cell wall is $1.5$ and the radius of all of the fillets is $0.5$.  The domain $\Omega$ is symmetric about the planes $x_1=3.75$, $x_2=3.75$, $x_3=10.75$, and $x_1=x_2$.

\begin{figure}
\hspace{.5in}\includegraphics[height=2.2in]{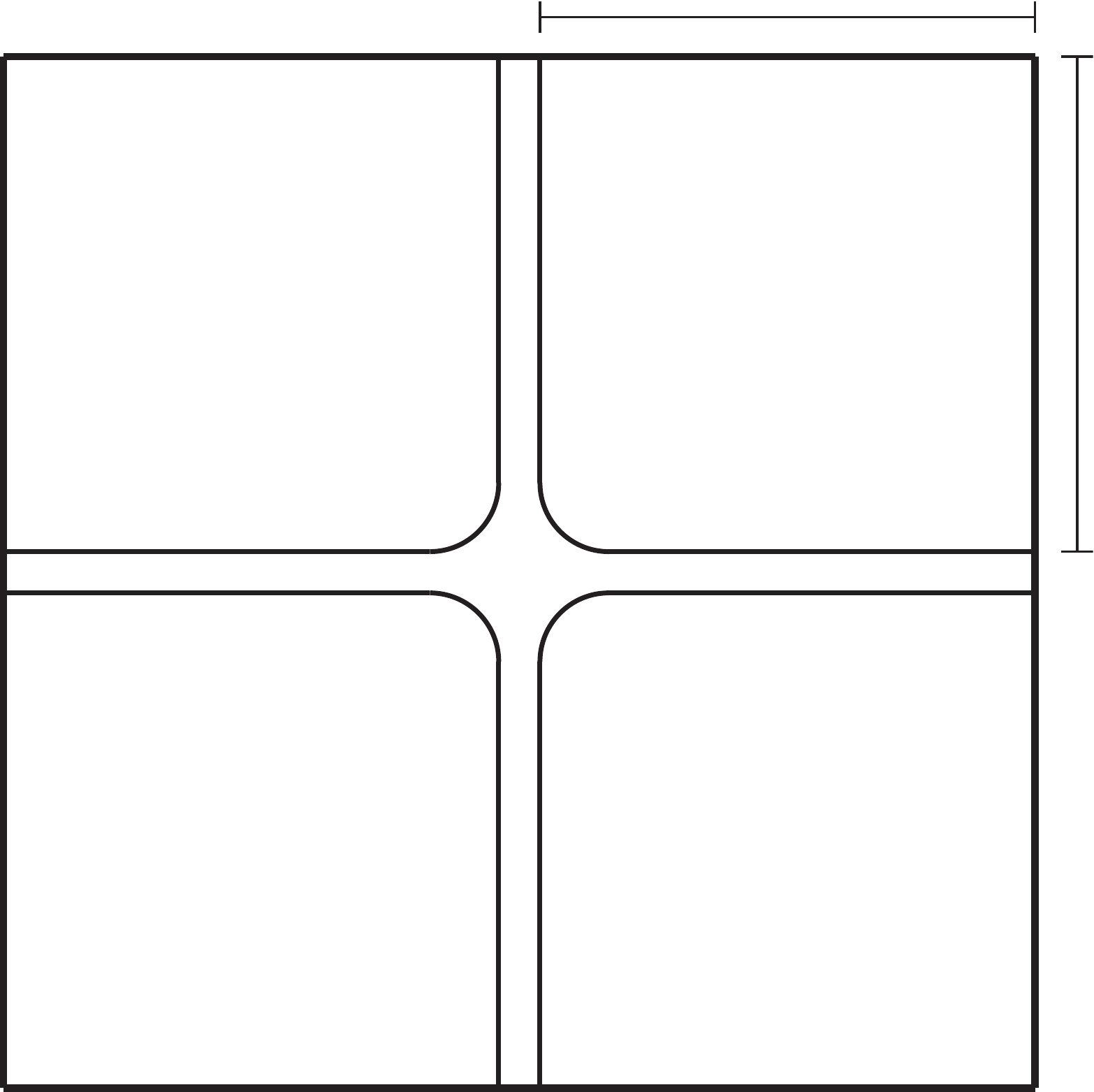}
\hspace{1.1in}\includegraphics[height=2.1in]{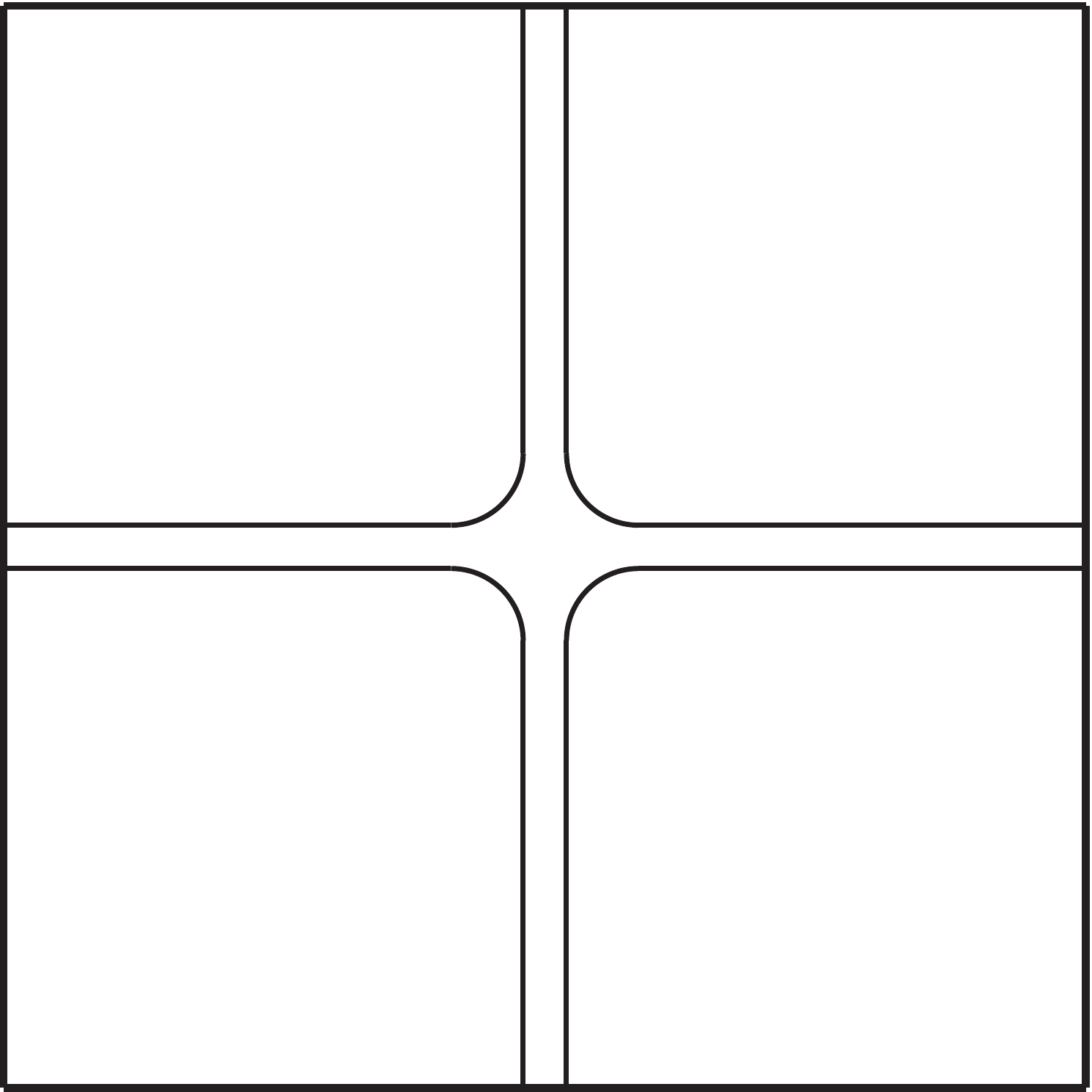}
\put(-285,110){\raisebox{.5pt}{\textcircled{\raisebox{-.9pt} {$1$}}}}
\put(-360,110){\raisebox{.5pt}{\textcircled{\raisebox{-.9pt} {$2$}}}}
\put(-360,35){\raisebox{.5pt}{\textcircled{\raisebox{-.9pt} {$3$}}}}
\put(-285,35){\raisebox{.5pt}{\textcircled{\raisebox{-.9pt} {$4$}}}}
\put(-290,160){$3.6\,\mu$m}
\put(-235,110){$3.6\,\mu$m}
\put(-45,110){\raisebox{.5pt}{\textcircled{\raisebox{-.9pt} {$5$}}}}
\put(-120,110){\raisebox{.5pt}{\textcircled{\raisebox{-.9pt} {$6$}}}}
\put(-120,35){\raisebox{.5pt}{\textcircled{\raisebox{-.9pt} {$7$}}}}
\put(-45,35){\raisebox{.5pt}{\textcircled{\raisebox{-.9pt} {$8$}}}}
\caption{A depiction of the subdomains within the central region of the domain.  The eight regions labeled $1$--$8$ are occupied by primary cell wall, and, hence, have cellulose microfibrils.  Separating these regions is the middle lamella which is $0.3\,\mu$m thick.  The regions with microfibrils are $3.6\,\mu$m on each side and are $0.6\,\mu$m thick.  The fillets have a radius of $0.5\,\mu$m.  The regions $1$--$4$ have larger $x_3$-values than regions $5$--$8$, and the $x_3$-axis goes through the lower left corner of regions $3$ and $7$.}
\label{XYcenter}
\end{figure}

The part of $\Omega$ contained within the box $(0,7.5)\times(0,7.5)\times(10,11.5)$ is called the central region and represents the upper and lower parts of the cell walls.  This region is divided into eight subdomains consisting of primary cell walls separated by a subdomain consisting of middle lamella.  The eight subdomains consisting of primary cell walls are labeled $1\mbox{--}8$ in Figure~\ref{XYcenter}.  The eight cells that make up $\Omega$ will be labeled according to which of these subregions they are in contact with.  The eight subdomains consisting of  cell walls are $3.6\,\mu$m in the $x_1$ and $x_2$-directions and $0.6\,\mu$m in the $x_3$-direction.  The subdomain consisting of middle lamella separating these eight regions is $0.3\,\mu$m thick.  To analyze the impact of the orientation of the microfibrils in the upper and lower regions  of the plant cell walls on the elastic deformation of a plant tissue we will consider different microfibrils orientations within the subdomains $1$--$8$.

\subsection{Governing equations and boundary conditions}\label{sectgebc}

The primary cell wall and the middle lamella are modeled as linearly elastic materials with different elastic properties. Let $\bbE$ be the elasticity tensor for the primary cell wall and middle lamella.  The value of $\bbE=\bbE(x)$ at any given point $x\in \Omega$ depends on whether that point lies in the middle lamella or in the primary cell wall.  Moreover, in the primary cell wall, the orientation of the cellulose microfibrils influences the elasticity tensor. This dependence will be specified in detail in the next subsection.

The boundary $\partial\Omega$ of the domain can be split into the union of three sets:
\begin{align}
\label{bdy0}\Gamma_0&=\{x\in \partial\Omega \ | \ x_1=0\ \text{or}\ x_2=0 \ \text{or}\ x_3=0\},\\
\label{bdymax}\Gamma_\text{max}&=\{x\in \partial\Omega \ |\ x_1=7.5\ \text{or}\ x_2=7.5 \ \text{or}\ x_3=21.5\},\\
\label{bdyI}\Gamma_I&=\partial\Omega\,\backslash\,(\Gamma_0\cup\Gamma_\text{max}).
\end{align}
The set $\Gamma_I$ is the part of $\partial\Omega$ in contact with the interior of the cells.  A pressure boundary condition corresponding to the turgor pressure will be imposed on $\Gamma_{I}$. On $\Gamma_\text{max}$ a tensile traction boundary condition will be specified.  Finally, $\Gamma_0$ is the part of the boundary of $\Omega$ that lies on the planes $x_1=0$, $x_2=0$, or $x_3=0$ associated with the planes of symmetry.  Thus, the displacement in the normal direction on $\Gamma_0$ must be zero.

Neglecting inertia and external body forces, the elasticity equation with these boundary conditions for the displacement $\bu$ is
\beqn\label{sysLE}
\begin{cases}
\text{div}(\bbE\be(\bu))=\textbf{0} & \text{in}\ \Omega,\\
\bu\cdot\bnu = 0 & \text{on}\ \Gamma_0,\\
(\bbE\be(\bu))\bnu\text{ is parallel to }\bnu   & \text{on}\ \Gamma_0,\\
(\bbE\be(\bu))\bnu = f\bnu & \text{on}\ \Gamma_\text{max},\\
(\bbE\be(\bu))\bnu = -p\bnu & \text{on}\ \Gamma_I,
\end{cases}
\eeqn
where $\be(\bu)=\frac{1}{2}(\nabla\bu+\nabla\bu^T)$ is the symmetric part of the gradient of the displacement and $\bnu$ is the exterior unit-normal to $\partial\Omega$.  A unique solution of \eqref{sysLE} exists in $H^1(\Omega,\Real^3)$ \cite{OSY} provided that $f \in L^2(\Gamma_\text{max})$, $p\in L^2(\Gamma_\text{I})$, and $\bbE$ satisfies the following conditions:
\begin{enumerate}
\item $|\bbE|$ is bounded in $L^\infty(\Omega)$.
\item There is a strictly positive $\alpha$ such that $\alpha|\bA|^2\leq\bA\cdot\bbE(x)\bA$ for all symmetric $\bA\in\Real^{3\times 3}$ and $x\in\Omega$.
\item $\bbE$ possesses major and minor symmetries, i.e.~$\bbE_{ijkl}=\bbE_{jikl}=\bbE_{klij}=\bbE_{ijlk}$.
\end{enumerate}

\subsection{The elasticity tensor}

Next, we specify the elasticity tensor $\bbE$ on the domain $\Omega$.  To do so, we must specify the elasticity tensor for the middle lamella and the primary cell wall for different microfibril configurations.  The macroscopic elastic properties of the primary cell wall are derived from the microscopic description of the elastic properties of the cell wall matrix and microfibrils using homogenization theory.  This requires the specification of the elastic properties of the cell wall matrix and the cellulose microfibrils.

%The elasticity tensor $\bbE$ varies depending on wether it is evaluated in the primary cell wall on the sides of the cells, in the primary cell wall within the center section, or the middle lamella.  In this section we specify the value of $\bbE$ in all of these regions.

The cell wall matrix is isotropic \cite{ZMSR}, and so the elasticity tensor of the matrix $\bbE_M$ is of the form
$$
\bbE_M\bA = 2\mu_M\bA + \lambda_M(\text{tr}\,\bA)\textbf{1},
$$
where the Lam\'e moduli $\mu_M$ and $\lambda_M$ are related to the Young's modulus $E_M$ and Poisson's ratio $\nu_M$ through
$$
E_M=\frac{\mu_M(2\mu_M+3\lambda_M)}{\mu_M+\lambda_M}\quad\text{and}\quad \nu_M=\frac{\lambda_M}{2(\mu_M+\lambda_M)}.
$$
We take $\nu_M=0.3$, which is common for biological materials, and $E_M=5$ MPa.  This value is lower than the Young's modulus measured for highly de-methylesterfied pectin gels considered in \cite{ZMSR} since the pectin within the cell wall matrix is not fully de-esterfied.

%
%which corresponds to a very elastic cell wall matrix that is expected in young plant cells.  
%
%
%%We assume that the middle lamella, with isotropic elasticity tensor $\bbE_{ML}$, has a Young's modulus of 15 MPa and Poisson's ratio of $0.3$.
%{\color{red} any citation here ?}

The cellulose microfibrils are not isotropic \cite{DMKM}, so we assume that they are transversely isotropic and, hence, the elasticity tensor $\bbE_F$ for the microfibrils is determined by specifying five parameters: the Young's modulus $E_F$ associated with the directions lying perpendicular to the microfibril, the Poisson's ratio $\nu_{F1}$ characterizing the transverse reduction of the plane perpendicular to the microfibril for stress lying in this plane, the ratio $n_F$ between $E_F$ and the Young's modulus associated with the direction of the axis of the microfibril, the Poisson's ratio $\nu_{F2}$ governing the reduction in the plane perpendicular to the microfibril for stress in the direction of the microfibril, and the shear modulus $Z_F$ for planes parallel to the microfibril.  A transversely isotropic elasticity tensor expressed in Voigt notation is given by
$$
\left (
\begin{matrix}
\alpha_2+\alpha_5 & \alpha_2-\alpha_5 & \alpha_3 & 0 & 0 & 0\\
\alpha_2-\alpha_5 & \alpha_2+\alpha_5 & \alpha_3 & 0 & 0 & 0\\
\alpha_3 & \alpha_3 & \alpha_1 & 0 & 0 & 0\\
0 & 0 & 0 & \alpha_4 & 0 & 0\\
0 & 0 & 0 & 0 & \alpha_4 & 0\\
0 & 0 & 0 & 0 & 0 & \alpha_5
\end{matrix}
\right),
$$
where $\alpha_i$, for $i=1,2,3,4,5$, are related to the five parameters described above through
\begin{align*}
\alpha_1&=\frac{E_F(1-\nu_{F1})}{n_F(1-\nu_{F1})-2\nu_{F2}^2},\\
\alpha_2&=\frac{E_Fn_F}{2n_F(1-\nu_{F1})-4\nu_{F2}^2},\\
\alpha_3&=\frac{E_F\nu_{F2}}{n_F(1-\nu_{F1})-2\nu_{F2}^2},\\
\alpha_4&=Z_F,\\
\alpha_5&=\frac{E_F}{2(1+\nu_{F1})}.
\end{align*}
We assign these parameters the values
$$
E_F = 15{,}000\, \text{MPa},\ \nu_{F1}= 0.3,\ n_F = 0.068,\ \nu_{F2}=0.06,\ Z_F = 84{,}842\, \text{MPa},
$$
which are chosen based on experimental results \cite{DMKM} and to ensure that the elasticity tensor for the microfibrils is positive definite \cite{Pad}.

We assume that the middle lamella is isotropic, with  elasticity tensor $\bbE_{ML}$, and has a Young's modulus of 15 MPa and Poisson's ratio of $0.3$.  It is know from experiments that the density of calcium-pectin cross-links strongly influence the elastic properties of the cell wall matrix and middle lamella \cite{WHH}. Thus, since in the middle lamella almost all pectin is de-esterified and the density of the pectin-calcium cross links is higher  than in the cell wall matrix, where usually  only $70\%$ of the pectin is de-esterified, we assume that the Young's modulus for the middle lamella is three times larger than the Young's modulus for the cell wall matrix.

The cellulose microfibrils are arranged periodically  within the cell wall matrix \cite{Tomasetal} and so standard techniques in homogenization theory, see e.g.~\cite{OSY}, yield a macroscopic elasticity tensor  for a plant cell wall from the microscopic description of the mechanical properties of a cell wall on the level of a single mibcrofibril.   In addition to the elastic properties of the microfibrils and cell wall matrix, the macroscopic elasticity tensor depends on the orientation of the cellulose mirofibrils. The components of this tensor are determined by solving unit cell problems, which have the form of the equations of linear elasticity and reflect the arrangement of the microfibrils.  

\begin{figure}
\hspace{.5in}\includegraphics[height=2.1in]{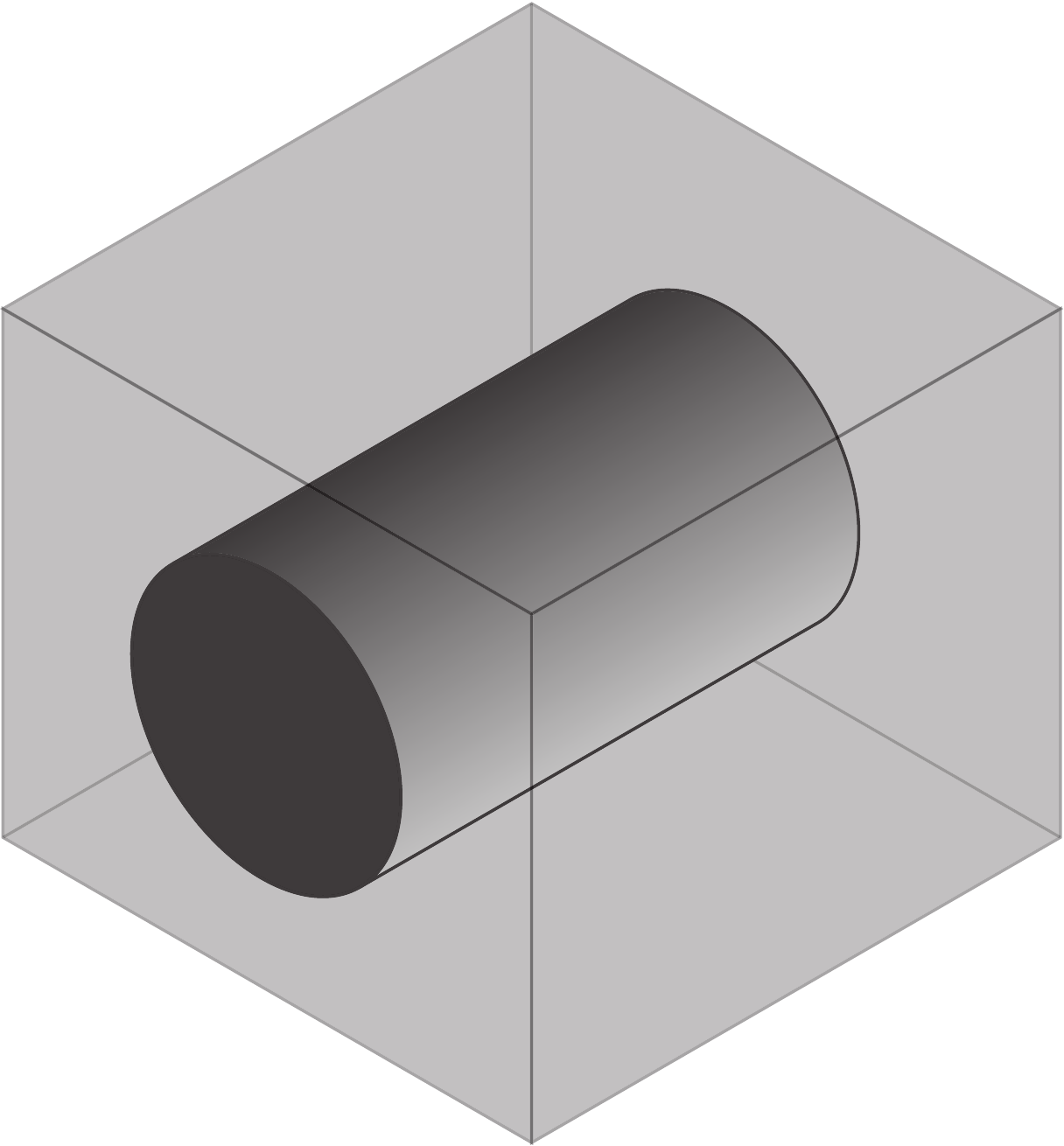}
\hspace{1.1in}\includegraphics[height=2.1in]{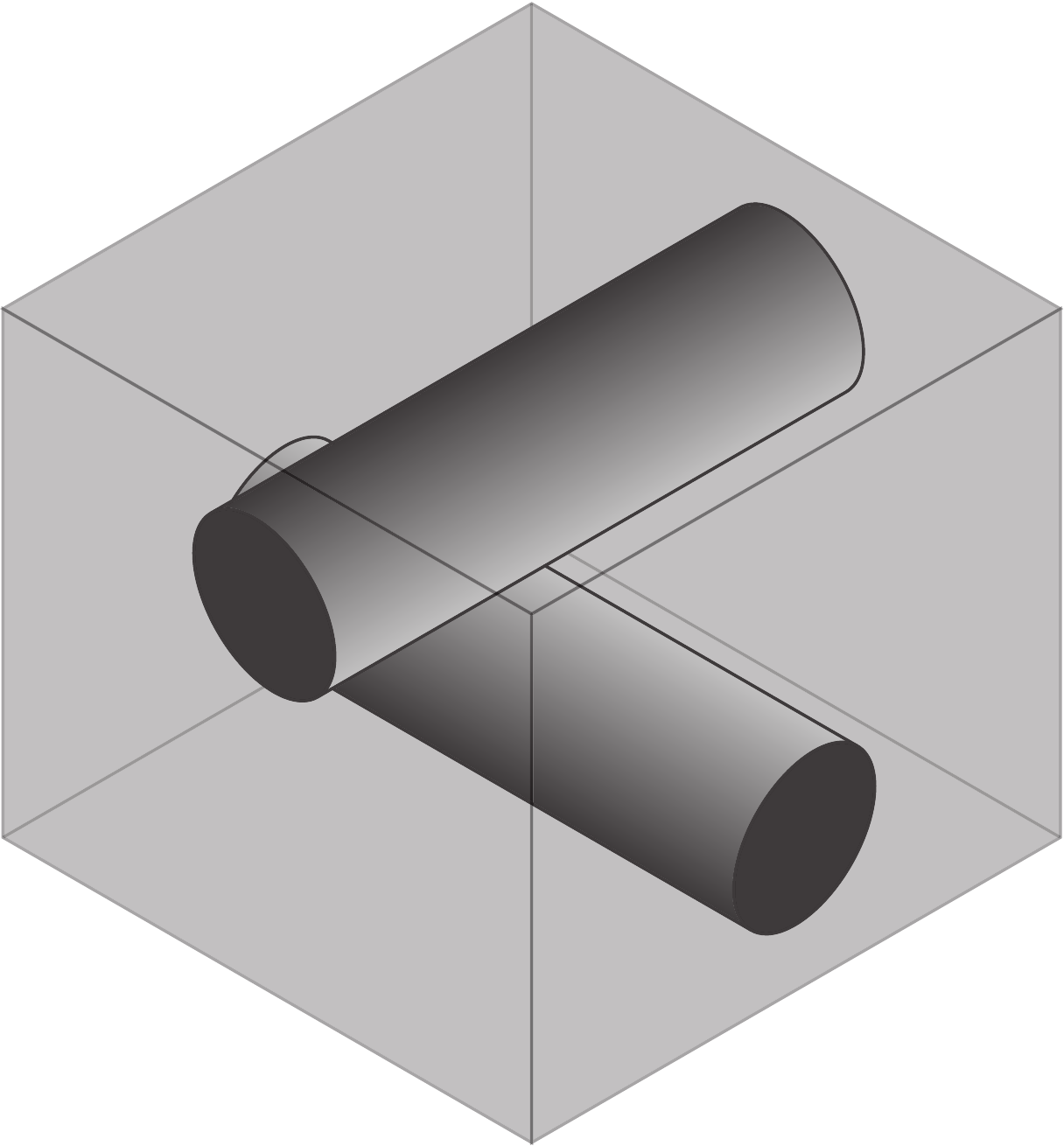}
\put(-301,-15){(a)}
\put(-304,9){\rotatebox[origin=c]{90}{$\vector(1,0){28}$}}
\put(-299,6.5){\rotatebox[origin=c]{30}{$\vector(1,0){28}$}}
\put(-323.5,-2){\rotatebox[origin=c]{150}{$\vector(1,0){28}$}}
\put(-290,18){$y_3$}
\put(-315,3){$y_2$}
\put(-280,3){$y_1$}
\put(-77,-15){(b)}
\put(-80.5,9){\rotatebox[origin=c]{90}{$\vector(1,0){28}$}}
\put(-75.5,6.5){\rotatebox[origin=c]{30}{$\vector(1,0){28}$}}
\put(-100,-2){\rotatebox[origin=c]{150}{$\vector(1,0){28}$}}
\put(-67,18){$y_3$}
\put(-92,3){$y_2$}
\put(-57,3){$y_1$}
\put(-77,-15){(b)}
\caption{A depiction of the unit cell $Y$ with two configurations of microfibrils.  Both unit cells are oriented so that the $y_3$-direction is pointing upward.  (a) A picture of the unit cell with one microfibril occupying the set specified in \eqref{YFone}.  (b) A picture of the unit cell with two microfibrils occupying the set specified in \eqref{YFtwo}.}
\label{fiborientations}
\end{figure}
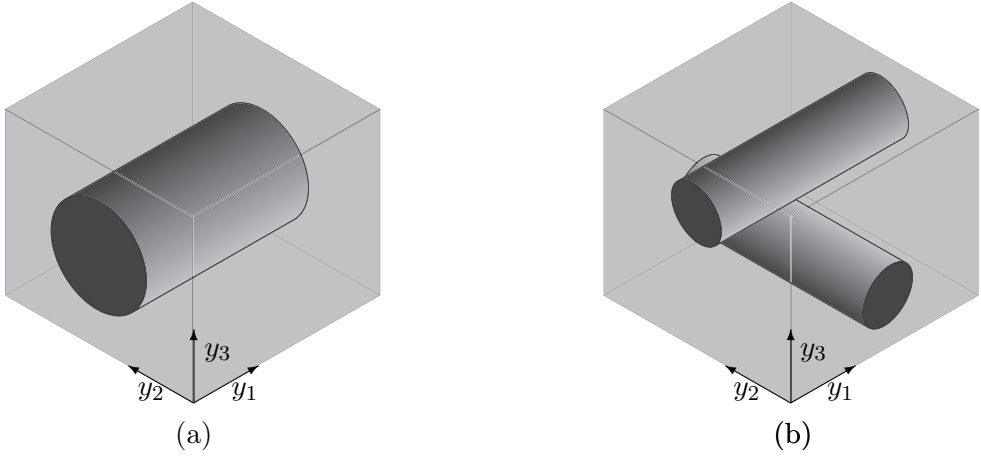

To specify the microstructure of a cell wall, consider the unit cell $Y=(0,1)^3$ and let $Y_M$ and $Y_F$ represent the parts of $Y$ occupied by the cell wall matrix and microfibrils, respectively, so that $Y_M$ and $Y_F$ are disjoint and $\overline Y = \overline Y_M\cup\overline Y_F$.  Two configurations of microfibrils within $Y$ are of primary interest.  The first is when there is only one microfibril in $Y$ occupying the set
\beqn\label{YFone}
Y_F=\{ y \in Y \ |\ (y_2-0.5)^2+(y_3-0.5)^2 < 0.25^2\},
\eeqn
and the other is when there are two microfibrils oriented in opposite directions  and occupy
\beqn\label{YFtwo}
Y_F = \{y\in Y \ |\ (y_2-0.5)^2+(y_3-0.75)^2<0.125^2\ \text{or}\  (y_1-0.5)^2+(y_3-0.25)^2<0.125^2\}, 
\eeqn
see Figure~\ref{fiborientations}.

Then, the elasticity tensor $\bbE_Y$ in $Y$ is given by
$$
\bbE_Y(y)=
\begin{cases}
\bbE_M & \text{if } y\in Y_M,\\
\bbE_F & \text{if } y\in Y_F,
\end{cases}
$$
and can be extended $Y$-periodically to all of $\Real^3$.  Consider a subdomain $U$ of $\Omega$ in which the cellulose microfibrils are arranged periodically with the orientation specified in $Y$ by \eqref{YFone} or \eqref{YFtwo}.  Let $\ve$ be a small parameter associated with the distance between the cellulose microfibrils.  
The  microfibrils of a plant cell wall are about $3$ nm in diameter and are separated by a distance of about $6$~nm, see e.g.~\cite{Colvin,KSJW,Thomas}, whereas  the thickness of a plant cell wall is of the order of a few micrometers.
To obtain the elasticity tensor for the part of the cell wall $U$ with a periodic microstructure  on the length scale of $\ve$ defined by the structure of $\ve Y$, the periodic extension of $\bbE_Y$ must be scaled appropriately.  Namely, the elasticity tensor in $U$ is given by
$$
\bbE^\ve(x)=\bbE_Y\left(\frac{x}{\ve}\right) \quad \rfa x\in U.
$$
Then homogenization theory yields a macroscopic elasticity tensor $\bbE_{\text{hom}}$ that describes a material whose behavior approximates the behavior of the  cell wall with elasticity tensor $\bbE^\ve$ when $\ve$ is very small \cite{OSY}.  In our situation $\ve\approx 10^{-3}$.  Moreover, $\bbE_\text{hom}$ is given by
\beqn\label{Ehom3D}
\bbE_{\text{hom},ijkl}=\dashint_Y[\bbE_{Y,ijkl}(y)+\bbE_{Y,ijpq}(y)\be_y(\bw^{kl})_{pq}(y)]dy,
\eeqn
where $\bw^{kl}\in H^1(Y,\Real^3)$ is the unique solution of %{\color{red} the condition \eqref{unitcell}$_2$ isn't the more precise formulation, but from a numerical point of view \eqref{unitcell}$_2$}
\beqn\label{unitcell}
\begin{cases}
\text{div}(\bbE_Y(\be(\bw^{kl})+\bb^{kl})) = \textbf{0}\qquad \text{in } Y,\\
\int_Y\bw^{kl}\, dy = \textbf{0},\ \bw^{kl}\text{ is } Y\text{-periodic},
\end{cases}
\eeqn
with $\bb^{kl}=\frac{1}{2}(\bb^k\otimes\bb^l+\bb^l\otimes\bb^k)$, where $(\bb^1,\bb^2,\bb^3)$ is the standard basis in $\Real^3$.  %It follows from the properties of $E_M$ and $E_Y$ that $\bbE_\text{hom}$ satisfies the conditions 1--3 mentioned at the end of Section~\ref{sectgebc}.

When $Y$ is given by \eqref{YFtwo}, the elasticity tensor given in \eqref{Ehom3D} will be denoted by $\bbE^{12}_\text{hom}$ as there are microfibrils in the $x_1$ and $x_2$-directions, while when $Y$ is given by \eqref{YFone} the elasticity tensor given in \eqref{Ehom3D} will be denoted by $\bbE^{1}_\text{hom}$ since the microfibrils are pointed in the $x_1$-direction.  Moreover, when $Y_F$ is given by \eqref{YFone}, then the microscopic elasticity tensor $\bbE^\ve$ depends only on two variables  $(x_2, x_3)$ and the unit cell problem \eqref{unitcell} can be reduced to a two-dimensional problem \cite{PS}.  To formulate this reduced problem, set $\hat Y=(0,1)^2$ and
$$
\hat Y_F = \{ (\hat y_2,\hat y_3)\in \hat Y \ |\ (\hat y_2-0.5)^2+(\hat y_3-0.5)^2 < 0.25^2\},
$$
so that $Y=(0,1)\times\hat Y$ and $Y_F=(0,1)\times\hat Y_F$.  It can be shown that
\beqn\label{Ehom2D}
\bbE_{\text{hom},ijkl}^1=\dashint_{\hat Y}[\bbE_{Y,ijkl}(0,\hat y)+\bbE_{Y,ijpq}(0,\hat y)\hat \be_{\hat y}(\hat \bw^{kl})_{pq}(\hat y)]d \hat y,
\eeqn
with $\hat \bw^{kl}\in H^1(\hat Y,\Real^3)$ being the unique solution of
\beqn\label{unitcell2D}
\begin{cases}
\hat{\text{div}}_{\hat y}(\bbE_Y(0,\hat y)(\hat\be(\bw^{kl})+\bb^{kl})) = \textbf{0}\qquad \text{in } \hat Y,\\
\int_{\hat Y}\hat\bw^{kl}\, d\hat y = \textbf{0},\ \hat\bw^{kl}\text{ is } \hat Y\text{-periodic},
\end{cases}
\eeqn
where for a function $\hat \bw\in H^1(\hat Y,\Real^3)$, the differential operators $\hat\be_{\hat y}$ and $\hat{\text{div}}_{\hat y}$ are defined by
$$
\hat\be(\hat\bw) = \left (
\begin{matrix}
0 & \frac{1}{2}\partial_{y_2} \hat\bw_1 & \frac{1}{2}\partial_{y_3} \hat\bw_1\\
\frac{1}{2}\partial_{y_2} \hat\bw_1 & \partial_{y_2} \hat\bw_2 & \frac{1}{2}(\partial_{y_2} \hat\bw_3+\partial_{y_3} \hat\bw_2)\\
\frac{1}{2}\partial_{y_3}\hat\bw_1 & \frac{1}{2}(\partial_{y_2} \hat\bw_3+\partial_{y_3} \hat\bw_2) & \partial_{y_3} \hat\bw_3
\end{matrix}
\right ) \text{ and } \hat{\text{div}}_{\hat y}\hat\bw = \partial_{y_2}\hat\bw_2+\partial_{y_3}\hat\bw_3, 
$$
see e.g. \cite{PS}. Reducing the unit cell problem to two dimensions allows for the consideration of a higher resolution mesh when solving the problem \eqref{unitcell2D}  numerically.

Besides considering the macroscopic elasticity tensor coming from microfibrils parallel to the $x_1$-axis, we will also consider the macroscopic elasticity tensor generated by microfibrils that are arranged in other directions in the $x_1x_2$-plane.  Given $\theta\in [-\pi/2, \pi/2]$, let $\bR^\theta$ denote the rotation about the $x_3$-axis through the angle $\theta$, so that
$$
\bR^\theta=\left (
\begin{matrix}
\cos\theta&\sin\theta&0\\
-\sin\theta&\cos\theta&0\\
0&0&1
\end{matrix}
\right ).
$$
The macroscopic elasticity tensor $\bbE^{1,\theta}_\text{hom}$ coming from a microstructure consisting of microfibrils aligned in the direction $\bR^\theta\bb^1$ is given by
$$
\bbE^{1,\theta}_{\text{hom},ijkl} = \bR^{\theta}_{ip}\bR^{\theta}_{jq}\bR^{\theta}_{kr}\bR^{\theta}_{ls}\bbE^1_{\text{hom},pqrs}.
$$
So, for example, the macroscopic elasticity tensor coming from a microstructure with microfibrils parallel to the $x_2$-axis is given by $\bbE^{1,\pi/2}_\text{hom}$.  

To summarize, the elasticity tensor $\bbE$ in the domain $\Omega$ is different in different regions within the primary cell wall.  In Figures~\ref{XYcrosssect} and \ref{XZcrosssect} we specify the regions  of the cell walls where the microfibrils are parallel to the $x_1$-axis,  i.e.\ $\bbE=\bbE^1_\text{hom}$, and  the regions of the primary cell wall where the microfibrils are parallel to the $x_2$-axis, i.e.\ $\bbE=\bbE^{1,\pi/2}_\text{hom}$.   Within subregion $i$, for $i=1,\dots,8$, of the central region, see Figure~\ref{XYcenter}, the elasticity tensor $\bbE$ will be set equal to $\bbE^i_\text{cen}$, where different choices of $\bbE^i_\text{cen}$ associated with different microfibril configurations will be considered.  Within the middle lamella, see Figures~\ref{XYcrosssect}--\ref{XYcenter}, there are no microfibrils and $\bbE=\bbE_{ML}$.  It follows from the properties of $\bbE_M$, $\bbE_F$, and $\bbE_{ML}$ that the macroscopic elasticity tensor $\bbE$ for the plant cell wall and middle lamella satisfies the conditions 1--3 mentioned at the end of Section~\ref{sectgebc}. Hence the problem \eqref{sysLE} describing macroscopic elastic properties of plant cell walls connected by middle lamella is well-posed. 

\section{Numerical results}\label{sectnum}

This section presents the results of the numerical simulations of the unit cell problems \eqref{unitcell} and \eqref{unitcell2D} necessary to calculate $\bbE^1_\text{hom}$ and $\bbE^{12}_\text{hom}$ and the simulations of the system \eqref{sysLE} for different  configurations of cellulose microfibrils  in the central region.  %To do this, the unit cell problems associated with $\bbE^1_\text{hom}$ and $\bbE^{12}_\text{hom}$ must be solved.%, which requires the specification of elasticity tensors $\bbE_M$ and $\bbE_F$.
All of the numerical simulations were done using FEniCS \cite{LMW,LW,OW}. This involved discretizing the domain using a nonuniform mesh and  applying  the continuous Galerkin method to solve the equations of linear elasticity. The resulting linear system was solved using the general minimal residual method with an algebraic multigrid preconditioner.

\subsection{Unit cell problems}\label{sectUCP}

It was observed experimentally that the calcium-pectin chemistry influences the mechanical properties of the cell wall matrix and middle lamella \cite{WHH}.  Hence in general, the elastic properties of the cell wall matrix depend on the density of the calcium-pectin cross-links $n$ and the microscopic elasticity tensor of the plant  cell wall  $\bbE^\ve$ is a function of $n$.   It was shown in \cite{PS} that  under the assumption of an isotropic cell wall matrix, the macroscopic elasticity tensor $\bbE_\text{hom}$ corresponding to any microfibril configuration is an affine function of the Young's modulus of the cell wall matrix.   From experiments \cite{ZMSR}, it is known that the Young's modulus $E_M$ of the cell wall matrix is a function of the density of calcium-pectin cross-links $n$ through the formula 
\begin{equation}\label{pectin}
E_M=0.775n+8.08,
\end{equation}
where $E_M$ has the units of MPa and $n$ has the units of $\mu$M.  Thus, knowing the macroscopic elasticity tensor  $\bbE_\text{hom}$  for two different values of $E_M$ determines the tensor for any value of $E_M$.   Then using \eqref{pectin} we obtain the macroscopic elasticity tensor for the cell wall for any calcium-pectin cross-links density $n$.  This approach enables us to analyse the changes in the mechanical properties of plant cell walls and tissues in response to the dynamics of calcium-pectin chemistry and changes in calcium-pectin cross-link density, which will be the subject  of future research. 

%Thus, instead of considering the effective elasticity tensor as a function of the Young's modulus of the cell wall matrix, we could consider it to be a function of the density of calcium-pectin cross-links, though this will not be done here.

%To obtain the macroscopic elasticity tensor for different values of $E_M$ and different cellulose microfibril orientations the unit cell problems are solved using FEniCS \cite{LMW,LW,OW}. This involves discretizing the domain using a nonuniform mesh with a higher density of vertices near the boundary between the cell wall matrix and the microfibrils and then discretizing the equations using continuous Galerkin elements. The resulting linear system is solved using the general minimal residual method with an algebraic multigrid preconditioner.

%number of vertices 18,645,460
%number of cells 37288870
For the numerical simulations to obtain the macroscopic elasticity tensor we consider two values for the Young's modulus: $E_M=10$ and $E_M=20$.  Then using the fact that  $\bbE_\text{hom}=\bbE_\text{hom}(E_M)$ is an affine function we obtain $\bbE_\text{hom}$ for $E_M=5$. To find $\bbE^1_\text{hom}$, the unit cell $\hat Y$ was discretized by a mesh  with $18{,}645{,}460$ vertices with a higher density of vertices near the boundary between the cell wall matrix and the microfibrils.  Using Voigt notation, the resulting effective elasticity tensor $\bbE^1_\text{hom}(E_M)$ for $E_M=10$ and $20$ are shown in Tables~\ref{bbE1E10} and \ref{bbE1E20}, respectively, to two decimal places.  Using the symmetry of the microstructure it can be shown analytically that the entries of the matrices $\bC^1(10)$ and $\bC^1(20)$ that are zero are exact and that some of the coefficients of the matrices $\bC^1(10)$ and $\bC^1(20)$ are equal \cite{PSsym}.  Specifically, for $E_M=10$ or $20$, $\bC^1(E_M)_{22}$ and $\bC^1(E_M)_{33}$ should be equal, $\bC^1(E_M)_{12}$ and $\bC^1(E_M)_{13}$ should be equal, and $\bC^1(E_M)_{55}$ and $\bC^1(E_M)_{66}$ should be equal. The largest scale involved in the numerical computations of the macroscopic elasticity tensors is determined by the Young's modulus of the microfibrils in the direction of the microfibrils and is equal to $220{,}000$ MPa. Using this scale, the relative error associated with $\bC^1(E_M)_{55}$ and $\bC^1(E)_{66}$ not being equal is on the order of $10^{-8}$.

\begin{table}
\begin{center}
$$\bC^1(10)=
\left (
\begin{matrix}
43333.24 & 12.51 & 12.51 & 0 & 0 & 0\\
12.51 & 19.27 & 7.59 & 0 & 0 & 0 \\
12.50 & 7.59 & 19.27 & 0 & 0 & 0 \\
0 & 0 & 0 & 5.34 & 0 & 0\\
0 & 0 & 0 & 0 & 9.30 & 0 \\
0 & 0 & 0 & 0 & 0 & 9.32 
%43367.2736 & 23.2326 & 23.2098 & 0.0241 & 0 & 0\\
%23.3214 & 21.0950 & 8.9765 & -0.0072 & 0 & 0\\
%23.1410 & 8.8423 & 21.1177 & -0.0005 & 0 & 0\\
%0.0179 & 0.0016 & 0.0095 & 5.6270 &  0 & 0\\
%0 & 0 & 0 & 0 & 13.9813 & -0.0005 \\
%0 & 0 & 0 & 0 & -0.0011 & 13.9893
\end{matrix}
\right)
$$
\end{center}
\vspace{-.1in}
\caption{The effective elasticity tensor $\bbE^1_\text{hom}$ expressed in Voigt notation to two decimal places when the Young's modulus of the matrix is $10$ MPa.}
\label{bbE1E10}
\end{table}

\begin{table}
\begin{center}
$$\bC^{1}(20)=
\left (
\begin{matrix}
43352.40 & 24.07 & 24.07 & 0 & 0 & 0 \\
24.07 & 37.75 & 14.89 & 0 & 0 & 0 \\
24.07 & 14.89 & 37.75 & 0 & 0 & 0 \\
0 & 0 & 0 & 10.44 & 0 & 0 \\
0 & 0 & 0 & 0 & 15.04 & 0 \\
0 & 0 & 0 & 0 & 0 & 15.05
%43400.1920 & 39.6508 & 39.6526 & -0.0040 & 0 & 0\\
%39.6065 & 39.6439 & 16.1447 & -0.0012 & 0 & 0\\
%39.6611 & 16.1635 & 39.6589 & 0.0001 & 0 & 0\\
%0.0130 & 0.0024 & 0.0057 & 10.7369 & 0 & 0\\
%0 & 0 & 0 & 0 & 19.7063 & 0.0004\\
%0 & 0 & 0 & 0 & -0.0014 & 19.7136
\end{matrix}
\right)
$$
\end{center}
\vspace{-.1in}
\caption{The effective elasticity tensor $\bbE^1_\text{hom}$ expressed in Voigt notation to two decimal places when the Young's modulus of the matrix is $20$ MPa.}
\label{bbE1E20}
\end{table}

%number of vertices 11750289
%number of cells 69492736

Similarly, discretizing $Y$ into a mesh with $11{,}750{,}289$ vertices, the calculated effective elasticity tensor $\bbE^{12}_\text{hom}(E_M)$ for $E_M=10$ is shown in Table~\ref{bbE12E10} and for $E_M=20$ is shown in Table~\ref{bbE12E20}, to two decimal places, using Voigt notation.  As before, it can be shown analytically that all of the entries of the matrices $\bC^{12}(10)$ and $\bC^{12}(20)$ that are zero are exact and that some of the components of the matrices $\bC^{12}(10)$ and $\bC^{12}(20)$ should be equal using the symmetry of the microstructure \cite{PSsym}.  Specifically, for $E_M=10$ or $20$, $\bC^{12}(E_M)_{11}$ and $\bC^{12}(E_M)_{22}$ should be equal, $\bC^{12}(E_M)_{13}$ and $\bC^{12}(E_M)_{23}$ should be equal, and $\bC^{12}(E_M)_{44}$ and $\bC^{12}(E_M)_{55}$ should be equal.  The largest relative difference between the components expected to be equal is of the order of $10^{-5}$.

\begin{table}
\begin{center}
$$\bC^{12}(10)=
\left (
\begin{matrix}
10927.86 & 99.60 & 67.85 & 0 & 0 & 0\\
99.60 & 10927.69 & 66.46 & 0 & 0 & 0\\
67.85 & 66.46 & 91.00 & 0 & 0 & 0\\
0 & 0 & 0 & 186.83 & 0 & 0\\
0 & 0 & 0 & 0 & 193.97 & 0\\
0 & 0 & 0 & 0 & 0 & 352.58 \\
\end{matrix}
\right)
$$
\end{center}
\vspace{-.1in}
\caption{The effective elasticity tensor $\bbE^{12}_\text{hom}$ expressed in Voigt notation to two decimal places when the Young's modulus of the matrix is $10$ MPa.}
\label{bbE12E10}
\end{table}

\begin{table}
\begin{center}
$$\bC^{12}(20)=
\left (
\begin{matrix}
10943.35 & 107.84 & 75.25 & 0 & 0 & 0\\
107.84 & 10943.18 & 73.87 & 0 & 0 & 0\\
75.25 & 73.87 & 106.55 & 0 & 0 & 0\\
0 & 0 & 0 &191.43 & 0 & 0\\
0 & 0 & 0 & 0 & 198.56 & 0\\
0 & 0 & 0 & 0 & 0 & 357.20\\
\end{matrix}
\right)
$$
\end{center}
\vspace{-.1in}
\caption{The effective elasticity tensor $\bbE^{12}_\text{hom}$ expressed in Voigt notation to two decimal places when the Young's modulus of the matrix is $20$ MPa.}
\label{bbE12E20}
\end{table}

The results of this section allow us to compute the elasticity tensor for any Young's modulus of the cell wall matrix, however in the following analysis  we only consider the case where $E_M=5$ MPa.

\subsection{Different boundary conditions and microfibril orientations in the central section}

%In Section~\ref{sectgeo} the subdomains of $\Omega$ in which the cellulose microfibrils are parallel to the $x_1$ or $x_2$-directions were specified.  

Using the numerical results for the effective elasticity tensor for different microfibril orientations, in this section we consider different microfibril orientations in the eight subregions of the central section, see Figure~\ref{XYcenter}, and different specifications of the turgor pressure $p$ and tensile force $f$ in problem \eqref{sysLE}.  %In all these cases, the Young's modulus of the cell wall matrix is $5$ MPa and the Young's modulus of the middle lamella is $15$ MPa.

We consider three different choices  for $p$ and $f$.  To describe these, set $p_\circ=209$ MPa, which is a common value for the turgor pressure \cite{BOB}.\\

\begin{description}

\item[(BC1)] Base case: $p=p_\circ$ and $f=5p_\circ$.
\item[(BC2)] No tensile tractions: $p=p_\circ$ and $f=0$.
\item[(BC3)] Different pressures, no tensile tractions: $p_1=p_4=p_5=p_8=p_\circ$ and $p_2=p_3=p_6=p_7=1.5p_\circ$, where $p_i$, for $i=1,\dots,8$, is the pressure in cell $i$,  and $f=0$.

\end{description}

\noindent For each of these boundary conditions we consider four different configurations of the microfibrils in the eight subregions of the center section.\\  %Let $\bbE^i_\text{cen}$, for $i=1,\dots,8$, denote the elasticity tensor for subdomain $i$ in the center section, see Figure~\ref{XYcenter}.
%\begin{itemize}

\begin{figure}
\hspace{.4in}\includegraphics[height=1.2in]{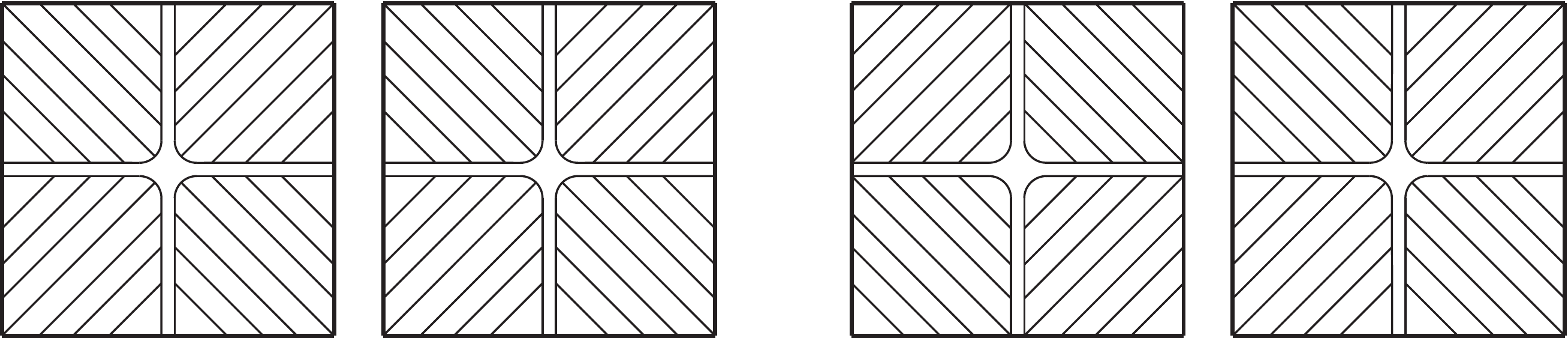}
\put(-343,62){\raisebox{.5pt}{\textcircled{\raisebox{-.9pt} {$1$}}}}
\put(-387,62){\raisebox{.5pt}{\textcircled{\raisebox{-.9pt} {$2$}}}}
\put(-387,18){\raisebox{.5pt}{\textcircled{\raisebox{-.9pt} {$3$}}}}
\put(-343,18){\raisebox{.5pt}{\textcircled{\raisebox{-.9pt} {$4$}}}}
\put(-245,62){\raisebox{.5pt}{\textcircled{\raisebox{-.9pt} {$5$}}}}
\put(-289,62){\raisebox{.5pt}{\textcircled{\raisebox{-.9pt} {$6$}}}}
\put(-289,18){\raisebox{.5pt}{\textcircled{\raisebox{-.9pt} {$7$}}}}
\put(-245,18){\raisebox{.5pt}{\textcircled{\raisebox{-.9pt} {$8$}}}}
\put(-27,62){\raisebox{.5pt}{\textcircled{\raisebox{-.9pt} {$5$}}}}
\put(-71,62){\raisebox{.5pt}{\textcircled{\raisebox{-.9pt} {$6$}}}}
\put(-71,18){\raisebox{.5pt}{\textcircled{\raisebox{-.9pt} {$7$}}}}
\put(-27,18){\raisebox{.5pt}{\textcircled{\raisebox{-.9pt} {$8$}}}}
\put(-124,62){\raisebox{.5pt}{\textcircled{\raisebox{-.9pt} {$1$}}}}
\put(-168,62){\raisebox{.5pt}{\textcircled{\raisebox{-.9pt} {$2$}}}}
\put(-168,18){\raisebox{.5pt}{\textcircled{\raisebox{-.9pt} {$3$}}}}
\put(-124,18){\raisebox{.5pt}{\textcircled{\raisebox{-.9pt} {$4$}}}}
\put(-318,-15){(a)}
\put(-99,-15){(b)}
\caption{(a) A depiction of the orientation of the cellulose microfibrils in (C1).  (b)  A depiction of the orientation of the cellulose microfibrils in (C2).}
\label{XYcenterDC}
\end{figure}
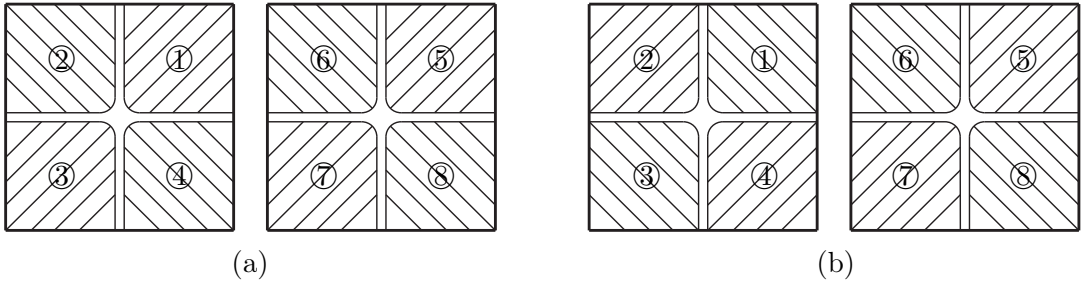

\begin{description}

\item[(C1)] In subregions 1, 3, 5, and 7 the microfibrils are parallel to $\bR^{\pi/4}\bb^1$ and in subregions 2, 4, 6, and 8 the microfibrils are parallel to $\bR^{-\pi/4}\bb^1$.Thus, for $i=1$, 3, 5, and 7, $\bbE^i_\text{cen}=\bbE^{1,\pi/4}_\text{hom}$, and for $i=2$, 4, 6, and 8, $\bbE^i_\text{cen}=\bbE^{1,-\pi/4}_\text{hom}$.  See Figure~\ref{XYcenterDC}(a).

\item[(C2)]  In subregions 2, 4, 5, and 7 the microfibrils are parallel to $\bR^{\pi/4}\bb^1$ and in subregions 1, 3, 6, and 8 the microfibrils are parallel to $\bR^{-\pi/4}\bb^1$.  Thus, for $i=2$, 4, 5, and 7, $\bbE^i_\text{cen}=\bbE^{1,\pi/4}_\text{hom}$, and for $i=1$, 3, 6, and 8, $\bbE^i_\text{cen}=\bbE^{1,-\pi/4}_\text{hom}$.  See Figure~\ref{XYcenterDC}(b).

\item[(C3)] In all of the eight subregions the orientation of the microfibrils on the microscale are generated by the unit cell depicted in Figure~\ref{fiborientations}(b).  Thus, $\bbE^i_\text{cen}=\bbE^{12}_\text{hom}$ for $i=1,\dots,8$.

\item[(C4)] There are no microfibrils in the center section.  Instead, the central section consists of middle lamella and, hence, $\bbE^i_\text{cen}=\bbE_{ML}$ for $i=1,\dots,8$.\\

\end{description}

The results of solving the system \eqref{sysLE} numerically using a mesh with $10{,}865{,}692$ vertices with a higher density of vertices in the central region with the boundary conditions (BC1) and (BC2) for the different configurations (C1)--(C4) are shown in Tables~\ref{tablemax1} and \ref{tablemax2}. For the boundary condition (BC3), the system \eqref{sysLE} was solved on a mesh with $3{,}801{,}841$ vertices with a higher density of vertices in the central region and the results are shown in Table~\ref{tablemax3}.  A lower resolution mesh was used for (BC3) because the iterative solver failed to converge in $10{,}000$ iterations when the higher resolution mesh was used.  For each combination of boundary conditions and microfibril configurations the maximal displacement in the positive and negative $x_1$, $x_2$, and $x_3$-directions are recorded to four significant figures. 

%For (BC1), the maximal displacements in the $x_1$ and $x_2$-directions occurred in the central region, while the maximal displacement in the $x_3$-direction occurred in the plane $x_3=21.5$.

\begin{figure}
\includegraphics[height=1.1in]{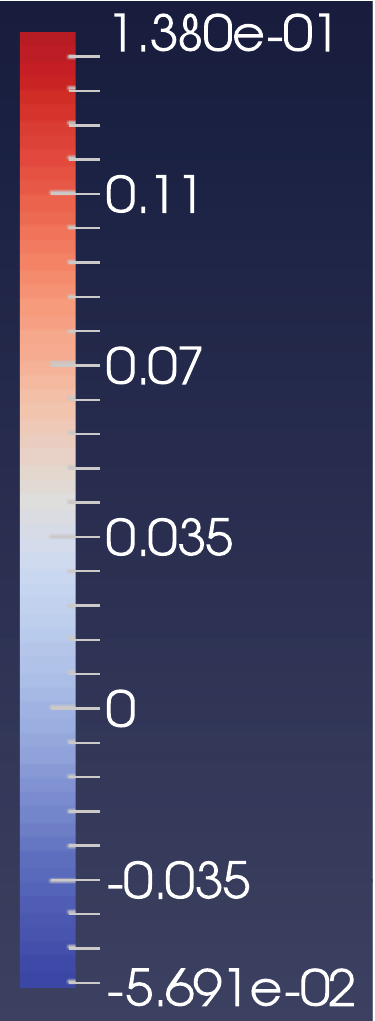}
\includegraphics[height=1.24in]{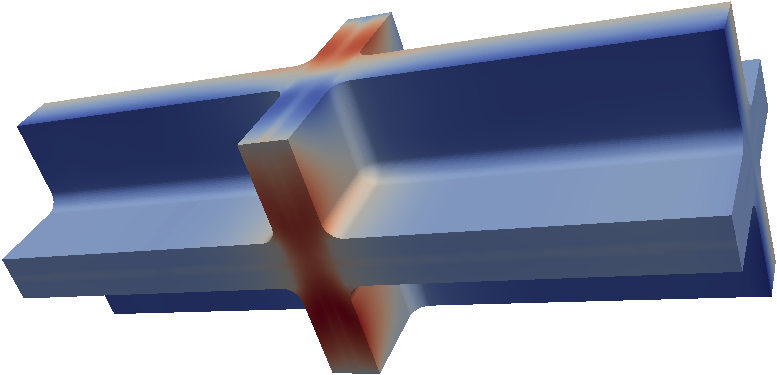}\quad 
\includegraphics[height=1.24in]{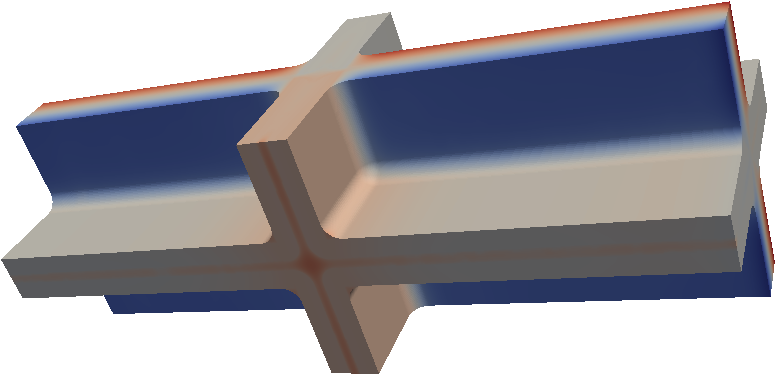}\; 
\includegraphics[height=1.13in]{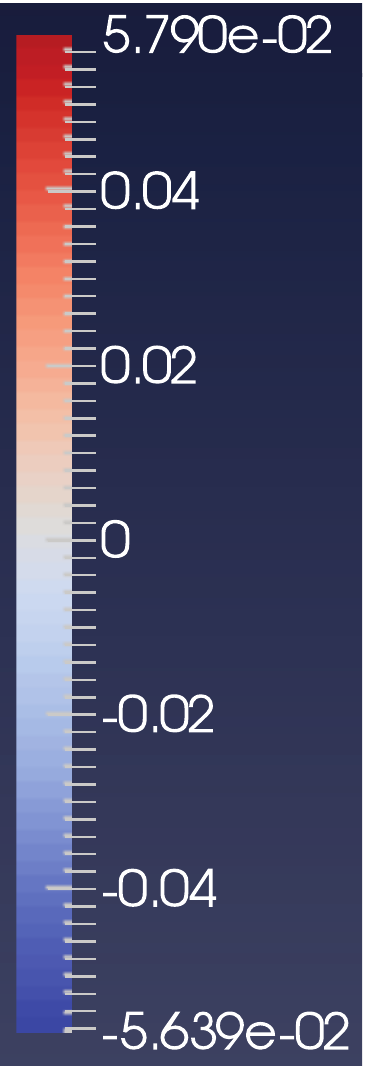}
\put(-320,-10){(a)}
\put(-120,-10){(b)}
\put(-209.2,78.2){\rotatebox[origin=c]{9}{$\vector(1,0){20}$}}
%\put(-123.2,103.2){\rotatebox[origin=c]{9}{$\vector(1,0){20}$}}
\put(-220.9,60.9){\rotatebox[origin=c]{229}{$\vector(1,0){20}$}}
%\put(-140.5,67.7){\rotatebox[origin=c]{229}{$\vector(1,0){20}$}}
\put(-226,78.15){\rotatebox[origin=c]{118}{$\vector(1,0){20}$}}
%\put(-215.5,51.15){\rotatebox[origin=c]{-62}{$\vector(1,0){20}$}}
\put(-190,74){$x_3$}
\put(-211,90){$x_1$}
\put(-227,69){$x_2$}
\caption{A depiction of the displacements in the $x_2$-direction for (BC1) with two different microfibril configurations: (a) configuration (C1) and (b) configuration (C3) dark red regions correspond to the areas of maximal displacement in the positive $x_2$-direction, while dark blue regions correspond to the areas of maximal displacement in the negative $x_2$-direction.  In these figures, the $x_1$-axis is pointed out of the page and upward and the $x_3$-axis is pointed to the right.}
\label{twodisp}
\end{figure}
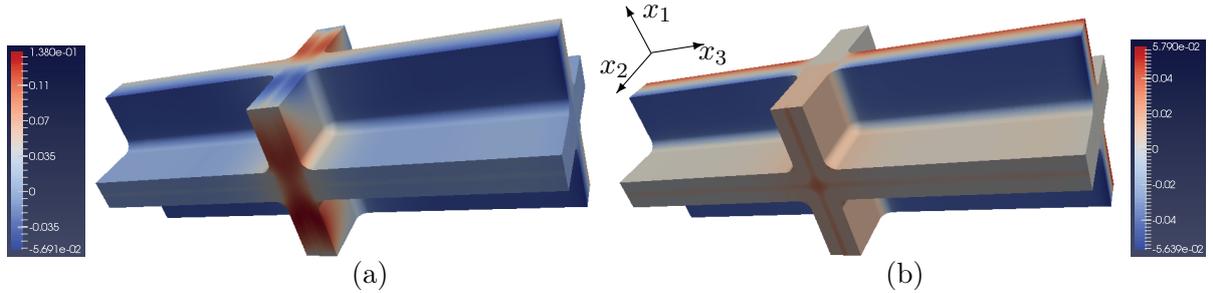

For configurations (C1), (C2), and (C4) and all boundary conditions the maximal displacements in the $x_1$ and $x_2$-directions occur within the central region, while for configuration (C3) the maximal displacements in these directions occur on the sides of the cell wall.  See Figure~\ref{twodisp}.  For the $x_3$-direction, the maximal displacement occurs at the plane $x_3=21.5$.

%8 cell mesh
%vertices 10865692
%cells 63289527

%8 cell rough mesh
%vertices 3801841
%cells 21663744

\begin{table}[h]
\begin{center}
\begin{tabular}{|c|c|c|c|c|c|c|}
\hline
(BC1) & \multicolumn{2}{c|}{$x_1$-direction} & \multicolumn{2}{c|}{$x_2$-direction} & \multicolumn{2}{c|}{$x_3$-direction} \\ \hline
 & negative & positive & negative & positive & negative & positive \\ \hline
(C1) & $-0.05804$ & $0.1377$ & $-0.05691$ & $0.1380$ & $-3.080\times 10^{-9}$ & $2.010$ \\ \hline
(C2) & $-0.05785$ & $0.2647$ & $-0.05696$ & $0.2659$ & $-8.263\times 10^{-9}$ & $2.009$ \\ \hline
(C3) & $-0.05765$ & $0.05756$ & $-0.05639$ & $0.05790$ & $-3.779\times 10^{-8}$ & $1.965$ \\ \hline
(C4) & $-0.05810$ & $0.2617$ & $-0.05682$ & $0.2617$ & $-4.496\times 10^{-9}$ & $1.981$ \\ \hline
\end{tabular}
\end{center}
\caption{The maximal displacement to four significant figures in the positive and negative $x_1$, $x_2$, and $x_3$-directions for boundary condition (BC1) and microfibril configurations (C1)--(C4).}
\label{tablemax1}
\end{table}

\begin{table}[h]
\begin{center}
\begin{tabular}{|c|c|c|c|c|c|c|}
\hline
(BC2) & \multicolumn{2}{c|}{$x_1$-direction} & \multicolumn{2}{c|}{$x_2$-direction} & \multicolumn{2}{c|}{$x_3$-direction} \\ \hline
 & negative & positive & negative & positive & negative & positive \\ \hline
(C1) & $-0.02037$ & $0.01509$ & $-0.01968$ & $0.01491$ & $-6.208\times 10^{-11}$ & $0.1183$ \\ \hline
(C2) & $-0.02034$ & $0.01451$ & $-0.01969$ & $0.01478$ & $-1.881\times 10^{-10}$ & $0.1184$ \\ \hline
(C3) & $-0.02035$ & $0.01425$ & $-0.01967$ & $0.01415$ & $-6.416\times 10^{-10}$ & $0.1209$ \\ \hline
(C4) & $-0.02031$ & $0.01640$ & $-0.01967$ & $0.01628$ & $-1.036\times 10^{-10}$ & $0.1212$ \\ \hline
\end{tabular}
\end{center}
\caption{The maximal displacement to four significant figures in the positive and negative $x_1$, $x_2$, and $x_3$-directions for boundary condition (BC2) and microfibril configurations (C1)--(C4).}
\label{tablemax2}
\end{table}

\begin{table}[h]
\begin{center}
\begin{tabular}{|c|c|c|c|c|c|c|}
\hline
(BC3) & \multicolumn{2}{c|}{$x_1$-direction} & \multicolumn{2}{c|}{$x_2$-direction} & \multicolumn{2}{c|}{$x_3$-direction} \\ \hline
 & negative & positive & negative & positive & negative & positive \\ \hline
(C1) & $-0.005638$ & $0.2041$ & $-0.05627$ & $0.03504$ & $-2.405\times 10^{-10}$ & $0.1449$ \\ \hline
(C2) & $-0.007259$ & $0.2017$ & $-0.05551$ & $0.03476$ & $-1.299\times 10^{-10}$ & $0.1451$ \\ \hline
(C3) & $-0.007191$ & $0.1990$ & $-0.05243$ & $0.03441$ & $-5.5574\times 10^{-11}$ & $0.1488$ \\ \hline
(C4) & $-0.005591$ & $0.2045$ & $-0.05604$ & $0.03508$ & $-2.789\times 10^{-10}$ & $0.1477$ \\ \hline
\end{tabular}
\end{center}
\caption{The maximal displacement to four significant figures in the positive and negative $x_1$, $x_2$, and $x_3$-directions for boundary condition (BC3) and microfibril configurations (C1)--(C4).}
\label{tablemax3}
\end{table}

\subsection{Smaller cells}\label{sectsmallcells}

%8 cell small mesh
%vertices 9490948
%cells 55351638

Besides considering the situations mentioned in the previous subsection, we also consider the case where the cells are smaller.  Namely, we consider cells $10$ $\mu$m smaller than those described in Section~\ref{sectgeo} so that the length of the domain in the $x_3$-direction is $11.5$ $\mu$m.  Looking at Figure~\ref{XZcrosssect}, this means that the $10$ $\mu$m measurements are decreased to $5$ $\mu$m.  Moreover, the boundaries \eqref{bdy0} and \eqref{bdymax} must be replaced with
\begin{align}
\label{bdy0sm}\Gamma_0&=\{x\in \partial\Omega \ | \ x_1=0\ \text{or}\ x_2=0 \ \text{or}\ x_3=5\},\\
\label{bdymaxsm}\Gamma_\text{max}&=\{x\in \partial\Omega \ |\ x_1=7.5\ \text{or}\ x_2=7.5 \ \text{or}\ x_3=16.5\}.
\end{align}
For the case of smaller cells, which will be referred to as (SM), we consider boundary condition (BC1) and configurations (C1)--(C4).  The results of solving the system \eqref{sysLE} numerically using a mesh with $9{,}490{,}948$ vertices with a higher density of vertices in the central region are shown in Table~\ref{tablemax4}.

\begin{table}[h]
\begin{center}
\begin{tabular}{|c|c|c|c|c|c|c|}
\hline
(SM) & \multicolumn{2}{c|}{$x_1$-direction} & \multicolumn{2}{c|}{$x_2$-direction} & \multicolumn{2}{c|}{$x_3$-direction} \\ \hline
 & negative & positive & negative & positive & negative & positive \\ \hline
(C1) & $-0.06053$ & $0.1377$ & $-0.06043$ & $0.1379$ & $-3.583\times 10^{-9}$ & $1.066$ \\ \hline
(C2) & $-0.05811$ & $0.2665$ & $-0.05788$ & $0.2657$ & $-3.942\times 10^{-8}$ & $1.058$  \\ \hline
(C3) & $-0.05744$ & $0.05842$ & $-0.05692$ & $0.0559$ & $-7.333\times 10^{-9}$ & $1.011$  \\ \hline
(C4) & $-0.06285$ & $0.2619$ & $-0.06275$ & $0.2621$ & $-2.214\times 10^{-9}$ & $1.041$ \\ \hline
\end{tabular}
\end{center}
\caption{The maximal displacement to four significant figures in the positive and negative $x_1$, $x_2$, and $x_3$-directions for the case of smaller cells for microfibril configurations (C1)--(C4).}
\label{tablemax4}
\end{table}

\section{Discussion}\label{sectdisc}

The data in Tables~\ref{tablemax1}--\ref{tablemax4} tell us several things about the affect of the presence and orientation of the cellulose microfibrils in the central region, i.e.\  the upper and lower ends of cell walls.  First of all, they have little affect on the expansion of the cells in the $x_3$-direction, as can be seen from looking at the last two columns in these tables. The cell wall is able to expand more in the directions perpendicular to the directions of the microfibrils since the microfibrils are much stiffer than the cell wall matrix and middle lamella.  Thus, changing the microfibril orientation within the $x_1x_2$-plane has little affect on the displacement in the $x_3$-direction.
 However, the expansion in the $x_1$ and $x_2$-directions are affected.  In particular, for (BC1) and (SM) when the microfibrils are arranged in the configuration (C3), the displacement in the positive $x_1$ and $x_2$-directions are $2/5$ of those for configuration (C1) and $1/5$ of those for configurations (C2) and (C4).  In configuration (C3) there are microfibrils oriented in both the $x_1$ and $x_2$-directions within the central region and it is expected that  for this configuration there would be less expansion in both directions. 
 The difference in the maximal deformations for different microfibril configurations in the central region  is less noticeable for the (BC2) and (BC3) boundary conditions. 
 Hence these results indicate that the orientation of microfibrils in the central section has an important impact on the deformation of plant cell walls and tissues in the case of  tensile traction boundary conditions. 

Comparing Tables~\ref{tablemax1} and \ref{tablemax2} we see that the presence of the tensile traction boundary condition causes the displacements in the positive directions to increase by an order of magnitude.  This is not surprising as the presence of more forces  causes larger displacements.  These tables also show that in the absence of a tensile traction boundary condition, the case (BC2), the displacement in the negative $x_1$ and $x_2$-directions is greater than the displacement in the positive $x_1$ and $x_2$-directions.  This is not the case for (BC1) and (SM) because of the applied tensile forces in the positive directions.  For (BC2), the reason that the absolute value of the displacement in the negative directions are greater than the maximal displacements in the positive $x_1$ and $x_2$-directions is the result of several factors.  Let us focus our discussion on the $x_1$-direction as the $x_2$-direction is similar.  The first thing we notice is that there are microfibrils in the $x_1$-direction, so any displacement in this direction is greatly hindered.  Next, we consider the boundary conditions on $\Gamma_I$ and $\Gamma_\text{max}$ and the fact that the turgor pressure $p$ is balanced with respect to the positive and negative directions.  The surfaces where the pressure is pointing in the positive $x_1$-direction are closer to a surface in which a no displacement in the $x_1$-direction boundary condition is imposed than the surfaces where the pressure is pointing in the negative $x_1$-direction.  Thus, the no displacement in the $x_1$-direction hinders the displacement in the positive $x_1$-direction more than the displacement in the negative $x_1$-direction.  This reasoning cannot be applied to the $x_3$-direction because there are no microfibrils pointing in $x_3$-direction.

%{\color{red} interpret table 3 (compare with table 2): displacement in different cells, more displacement in the $x_1$-direction because that is the direction of the gradient of the pressure, more displacement in the $x_3$-direction than in (BC1) case, }

Comparing Table~\ref{tablemax2} and Table~\ref{tablemax3} we can see the effect of increasing the pressures in some of the cells.  First, notice that the displacements in the positive directions are larger in the case  (BC3) than in the case (BC2).  This is because in (BC3) the pressure in cells 2, 3, 6, and 7 is greater than in the (BC2) case.  Also notice that in the  case (BC3) the displacement in the positive $x_1$-direction is greater than the displacement in the positive $x_2$ and $x_3$-directions, which is caused by the the position of the cells with the larger pressure.  Namely, there is a pressure difference between the cells that are aligned in the $x_1$-direction. The impact of the presence and orientation of microfibrils   on  the displacement  is relatively small. 

Finally, looking at Tables~\ref{tablemax1} and \ref{tablemax4}, one sees that the only significant difference in the case of small cells is in the displacement in the $x_3$-direction---there is twice the displacement for the larger cells than for the smaller cells.  This is in accord with Hooke's law, which tells us that the elongation of an elastic bar under an applied load is a linear function of the length of the bar.

To conclude, the orientation of the microfibrils in the upper and lower parts of plant cell walls have no effect on the elongation of the cells, but will influence their radial expansion and growth. 
%However, assuming that on the sides of the cell wall microfibrils are arranged in rings around the cells and considering the same turgor pressure in all cells and/or the same  tensile traction boundary conditions, the expansion in the radial direction is one order of magnitude lower than the expansion in the direction of the elongation. 
 The results in Table~\ref{tablemax3} show that  different pressures in neighbouring cells, which can be observed during the growth process,  influence the direction of the maximal displacement (here the maximal displacement in the $x_1$-direction is due to pressure distributions).  It follows from our results that only in the case of  directed tensile forces applied to  plant cells and tissues will  the orientation of the microfibrils in the lower and upper parts of cell walls play a role. Hence, for cells  where the main acting forces are turgor pressure, the orientation of the microfibrils in the lower and upper parts of the cell walls is not essential and cells will choose the most energy efficient way to orient the microfibrils in these parts of the cell walls. However in the parts of the tissues where there is a strong directed tissue tension, the importance of the orientation of the microfibrils in the upper and lower parts of the cell walls may be important. In our studies we assumed that the microfibrils on the sides of the cell wall are arranged in fixed rings around the cells  without considering possible  sliding of the microfibrils during the expansion. The affect of the sliding of the microfibrils on the deformation of plant cells and tissues in combination with different arrangements of microfibrils in the upper and lower parts of the cell walls  will be the subject of future studies.

%\begin{acknowledgements}
%\end{acknowledgements}

%\bibliographystyle{spbasic}      % basic style, author-year citation
\bibliographystyle{acm}      % mathematics and physical sciences
\bibliography{8cell} 

%% Non-BibTeX users please use
%\begin{thebibliography}{}
%%
%% and use \bibitem to create references. Consult the Instructions
%% for authors for reference list style.
%\end{thebibliography}

\end{document}